%% file: main.tex
\documentclass[letterpaper]{article}
\input{packages}
\usepackage{arxiv}
\usepackage{helvet} 
\usepackage{courier} 
\usepackage[utf8]{inputenc} 
\usepackage[T1]{fontenc}    
\usepackage{hyperref}       
\usepackage{url}            
\usepackage{booktabs}       
\usepackage{amsfonts}       
\usepackage{nicefrac}       
\usepackage{microtype}      
\usepackage{cleveref}       
\usepackage{lipsum}         
\usepackage{graphicx}
\usepackage{doi}
\usepackage{tikz}

\title{Multimedia and Visual Analytics in the Agentic Era}

\usepackage{authblk}

\newbox{\orcid}\sbox{\orcid}{\includegraphics[scale=0.06]{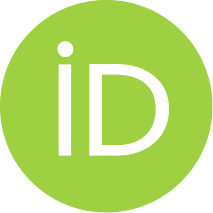}} 
\author[1]{%
	\href{https://orcid.org/0000-0003-4097-4136}{\usebox{\orcid}\hspace{1mm}Marcel  Worring\thanks{\texttt{m.worring@uva.nl}}}%
}
\author[2]{%
	\href{https://orcid.org/0000-0002-6743-3607}{\usebox{\orcid}\hspace{1mm}Jan Zah{\'a}lka\thanks{\texttt{jan.zahalka@cvut.cz}}}%
}
\author[3]{
	\href{https://orcid.org/0000-0002-6743-3607}{\usebox{\orcid}\hspace{1mm}Stef van den Elzen \thanks{\texttt{s.j.v.d.elzen@tue.nl}}}%
} 
\author[4]{
	\href{https://orcid.org/0000-0001-8076-1376}{\\ \usebox{\orcid}\hspace{1mm} Maximilian T. Fischer \thanks{\texttt{max.fischer@uni-konstanz.de}}}%
}
\author[5]{
	\href{https://orcid.org/0000-0001-7966-9740}{\usebox{\orcid}\hspace{1mm}Daniel A. Keim \thanks{\texttt{keim@uni-konstanz.de}}}%
}
\affil[1]{University of Amsterdam}
\affil[2]{Czech Technical University in Prague}
\affil[3]{Eindhoven University of Technology}
\affil[4,5]{University of Konstanz}

\renewcommand{\shorttitle}{\textit{Multimedia and Visual Analytics in the Agentic Era}}
\graphicspath{{figs/}{figures/}{pictures/}{images/}{./}} 

\input{commands}
\begin{document}

\maketitle

\begin{abstract}
Professional users need tools to help them gain actionable insights from large multimedia collections. Foundation models and AI agents have rapidly changed the playing field, and improving their accuracy, trustworthiness, and reasoning capabilities are active topics in the computer vision, machine learning, and multimedia communities. Most current research focuses on benchmark driven algorithmic improvements. The multimedia community is the place to go beyond algorithms and consider complete multimedia analytics systems that support professional users in their complex tasks and achieve a true teaming of humans and AI. Supporting users with machine learning and visualizations has been studied for decades in the visual analytics field. In this paper, we propose a framework to bring multimedia and visual analytics together and indicate how it could impact current and new multimedia analytics solutions. Additional information can be found at \url{https://staff.fnwi.uva.nl/m.worring/analytics-model.html}
\end{abstract}

\keywords{Multimedia Analytics, Foundation Models, Human-AI Teaming, Visual Analytics Agents, Human-in-the-Loop}


\begin{figure}
  \includegraphics[width=\textwidth]{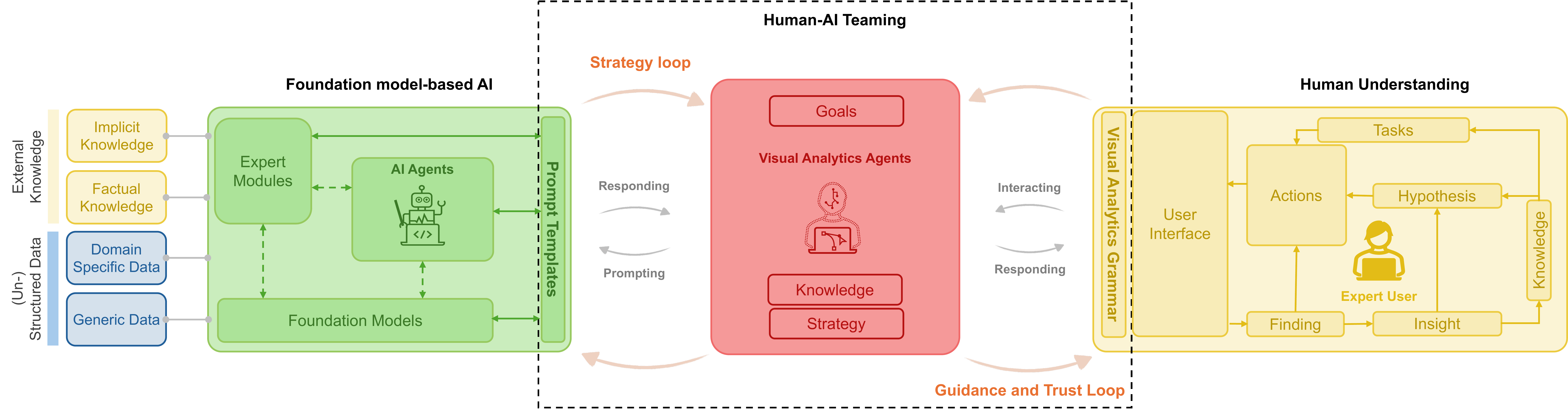}
  \caption{%
  Overview of our proposed framework for multimedia analytics using \bgcolorbox[ColorAI!30]{foundation models and agentic AI} and explaining how they can be combined with\bgcolorbox[ColorHumanUnderstanding!30]{human understanding}. Central to the model is a formalization of \bgcolorbox[ColorHumanAITeaming!30]{Human-AI Teaming} through \textcolor{ColorVAAgent!40!black}{visual analytics agents} that mediate between \textcolor{ColorExpert!60!black}{expert users} and \textcolor{ColorAIAgent!60!black}{AI} through numerous \bgcolorboxstriped[ColorChannel!25]{communication channels}. These \textcolor{ColorVAAgent!40!black}{agents} in a \textcolor{ColorAction!75!black}{strategy loop} with the AI models optimize their internal strategy through \textcolor{ColorAIAgent!60!black}{template based prompt abstraction}, while in the \textcolor{ColorAction!75!black}{guidance and trust loop} they aim to optimally support the user's task through jointly addressing the problem at hand based on a \textcolor{ColorExpert!60!black}{visual analytics grammar}. 
  }
  \label{fig:teaser}
\end{figure}


\section{Introduction}
Today, multimodal information sources have revolutionized work practices in law enforcement~\cite{FiHiJe2022VAIntelligence,gisolf2025hypergraphs}, journalism~\cite{fu2023more}, marketing~\cite{Rietveld2020Instagram}, finance~\cite{Zhu_multimodal_finance2025}, medical imaging~\cite{Bannach2017radiomics}, cultural heritage~\cite{Bin_GalleryGPT_2025,Rachabatuni_artChatBot_2024,Shuai_artRag_2025}, and climate research~\cite{Kokoshka2014climate}.
\textbf{Foundation models} (FMs) including large language models (LLMs) and multimodal LLMs (MLLMs) brought surprising jumps in the accuracy of Natural Language Processing (NLP),  automatic speech recognition, and text-to-speech, as well as \textit{vision} tasks, such as semantic understanding, image classification, and object recognition. Foundation models now also have the generative ability to create text, data visualizations, images, videos, and software code. AI in notoriously difficult but structured tasks, such as playing Chess and Go or predicting protein structure outperform humans. In unstructured domains foundation models still have several limitations; human intuition, creativity, and analytic capabilities are essential and  still outperform many AI solutions~\cite{MorrisAGI2024}.

As argued in the AI~\cite{mosqueira2023human}, visual analytics~\cite{keim2008visual, sacha2016human}, and the human-computer interaction (HCI) \cite{Wang2020} communities, \textbf{human-machine collaboration} is the best way forward.
As a prominent example, conceptual models that bring together users and systems have been studied in the \emph{visual analytics} (VA) community for decades, for example through Keim's high-level VA model~\cite{keim2008visual}, the Knowledge Generation Model~\cite{sacha2014knowlegde} by Sacha et al. integrating cognitive processes, or the exploration of guided VA~\cite{perez2022typology}.
Existing VA models often inherently assume specialized analysis components and a user that controls the steering and analysis process through input based on visualizations and programmatic guidance.
This modular analytics and user-in-control paradigm does not adequately capture the emerging role of FMs as collaborative agents~\cite{Hutchinson.FoundationalVAOpportunities.2025}. The combination of multimedia analysis and visual analytics has led to the field of \textbf{multimedia analytics} (MMA)~\cite{chinchor2010multimedia,zahalka2014towards}.

With the rapidly growing capabilities of foundation models, multimedia analytics systems are now being built directly on top of these rather than by using analytical libraries and adaptable algorithms. This approach is suitable for quick prototyping, but introduces a set of functional and strategic challenges. Functional challenges are stemming from the fact that the primary purpose of FMs is \emph{not} analytics. From the strategic perspective, direct FM usage in analytics often requires the largest available FMs to achieve sufficient analytical quality and understanding, which might be unavailable due to cost, security and/or vendor lock-in reasons. Marrying the power of FMs with a ``visual analytics first'' framework unlocks the potential of using smaller, open-source models, in turn increasing the adoption of FM-powered multimedia analytics.

Next to the multimedia analysis capabilities, we observe the increasing shift from the (still common) text-based prompt dialogue~\cite{schulhoff2025promptreportsystematicsurvey}, reminiscent of the famous Eliza system of 1966~\cite{weizenbaum1966eliza}, to a more integrated, \textbf{contextual agentic style communication} with tool-use, using standards like the Model Context Protocol (MCP)~\cite{mcp} and observation-based agents. Integrating FMs and agentic workflows often goes beyond replacing analytical modules, evolving from passive support tools to collaborative, equal partners~\cite{Hutchinson.FoundationalVAOpportunities.2025} integrated into a single MMA system. 
Human-AI Teaming from an HCI perspective requires mutual goal understanding, task co-management, and shared progress tracking~\cite{Wang2020}.
Current models for human-AI teaming remain on an abstract, conceptual level~\cite{Hutchinson.FoundationalVAOpportunities.2025, Bernard.HumanDataModelVA.2025}.

It remains an open question how high-level concepts in current models can be described technically in an agentic VA context, how actions and interactions differ, and how actual knowledge gains are realized~\cite{Bernard.HumanDataModelVA.2025}. Solutions to support professionals must not only consider how to reveal the implicit information present in datasets, but also how to optimally align analysis tools with the analytic mental process of the domain expert.
Based on the abstract conceptual ideas and existing models, we distill an \textbf{agentic framework for multimedia analytics}, introducing and formalizing VA agents that work alongside AI agents, and define the  data, visualization, and interaction channels between models, UI, and humans more explicit than in previous work. We make the following \textbf{contributions}:
\begin{itemize}
    \item A  new agentic \textbf{multimedia analytics framework}  (\autoref{fig:teaser}) integrating foundation models and agentic approaches for modern system design, extending established frameworks and ideas.
    \item A formalization of \textbf{communication channels} (data, visualization, interaction) in visual analytics to support human-AI teaming (\autoref{fig:human-in-the-loop}). 
    \item An \textbf{explicit pipeline} for user interaction processing that harnesses the full power of foundation models, while providing guidance and aiding the buildup of user trust in AI.
\end{itemize}
In the remainder of the paper, after reviewing related work (\autoref{sec:related-work}), we elaborate on the framework design (\autoref{sec:framework_construction}), including our framework construction process. We then look at foundation model based AI (\autoref{sec:AImodels}) after which we consider the challenges from the human understanding side (\autoref{sec:human_analytics}). We then consider human-AI teaming (\autoref{sec:human-ai-teaming}) based on visual analytics agents. Finally, we consider how to assess methods (\autoref{sec:evaluation}) developed using the framework and discuss the impact the framework (\autoref{sec:conclusion}) could have on multimedia applications. 

\section{Related work}

\label{sec:related-work}


\parheader{Interactive and Agentic Machine Learning} Classic interactive multimodal learning (IML) methods, relevance feedback and active learning, have long been central to multimedia analytics~\cite{jiang2019recent,khan2020interactive,mosqueira2023human,zahalka2017blackthorn}. Combined with semantic descriptors, IML has powered numerous multimedia analytics~\cite{batch2023uxsense, he2023videopro,worring2016pivot} and retrieval systems~\cite{khan2024exquisitor,Lokuc2023VBS, zahalka2015interactive}. 
Foundation models have fundamentally changed how users interact with analytical systems. Where IML required carefully designed feature spaces and task-specific training, FMs enable interaction through natural language prompts and in-context learning, replacing dedicated models with inference-time adaptation~\cite{schulhoff2025promptreportsystematicsurvey}. This shifts the interaction from example labeling to high-bandwidth, mixed-media dialogue where the user specifies intent and the model handles execution [33]. The move to agentic AI~\cite{Guo2024MultiAgentSurvey890,Xi2025LLMagentsurvey} introduces a further shift. Rather than responding to isolated prompts, AI agents autonomously decompose complex tasks, reason through multi-step strategies~\cite{wei2022chain,yao2023tree}, and use external tools via protocols such as MCP~\cite{mcp}. Agent architectures with planner, executor, and controller components~\cite{zhao2025lightweightVA} enable semi-autonomous analytical workflows. Multi-agent systems~\cite{Guo2024MultiAgentSurvey890,Sheng2026MultiAgent} further allow parallel, coordinated analysis. This evolution redefines the human-AI relationship from a user controlling algorithms to human-AI teaming.


\parheader{Visual Analytics Models} Keim et al.~\cite{keim2008visual,keim2010mastering} phrased a simple yet powerful abstraction model that combines data, models, visualizations, and humans.
Green et al.~\cite{Green.VAHumanCognition.2008, Green.HumanCognitionModelVA.2009} introduced cognition-based models, which Sacha~\cite{sacha2014knowlegde} then added as cognition aspects~\cite{Green.VAHumanCognition.2008} to Keim's VA model.
Viewing visual analytics as an interactive model building step was also echoed by Andrienko et al.
\cite{Andrienko.ModelBuilding.2018}.
In contrast to these technical and knowledge-centric models, design spaces like Munzner's nested model~\cite{munzner2015visualization} follow a task-based approach to visualization and VA, guiding exploration along the what?, why?, and how? questions, to work out technical and algorithmic requirements.
Missing formalization on guidance finally led to an extension to guided visual analytics~\cite{perez2022typology}, beginning to open up the silo-ed algorithmic modules, but is still limited in autonomy and suggestion capabilities.
The survey by Bernard provided an overview of the realization of the different VA models~\cite{Bernard.HumanDataModelVA.2025} and the benefits they offer humans from an external stakeholder perspective.
Existing visual analytics models have established a solid understanding of user needs and cognitive processes and have successfully aligned these insights with the then available machine learning capabilities. FMs are far more powerful and hence the models should be reconsidered.

\parheader{Integrated Models} Vats et al.~\cite{vats2024surveyhumanaiteaminglarge} survey a general-purpose wide-ranging view of how large FMs augment human-AI teaming across sectors, emphasizing ethical, safety, and trust considerations. Our work further narrows to MMA, introducing an agentic framework for multimodal mixed-initiative analysis. Hutchinson et al.~\cite{Hutchinson.FoundationalVAOpportunities.2025} examine FM integration into VA processes, identifying alignment, reasoning, and evaluation challenges. We aim to address these within a structured model, embedding the processes into a continuous human-AI collaboration loop with explicit interaction and VA agent roles. Similarly, Monadjemi et al.~\cite{monadjemi2023agentbased} and Bernard~\cite{Bernard.HumanDataModelVA.2025} formalize human, data, and model roles, but do not target FMs or multimodal integration. In contrast to Monadjemi et al.~\cite{monadjemi2023agentbased}, who view a VA agent as either a human or AI contributing to the analytic process, we design AI agents that have explicit knowledge of VA and play an important part in the human-AI teaming. 
System-level work, e.g., LightVA~\cite{zhao2025lightweightVA} implements LLM-based architectures for task-driven VA, overlapping with parts of our model. However, LightVA focuses on efficiency and LLMs, while we generalize to a broader range of VA scenarios. The Model Context Protocol~\cite{mcp} addresses assessment and interoperability but lacks VA-specific conceptualization. We aim to fill this gap by providing a model tailored to foundation-model agentic AI and VA, and provide guidance on applying these elements in multimedia analytics.

\parheader{Evaluation Models} Benchmark based evaluation of AI is well established, evaluation of human-AI teaming in multimedia analytics is not. 
As multimedia analytics focuses on flexible exploration rather than strict tasks,  operates on complex multimodal data, and requires human-AI collaboration, North's~\cite{north2006toward} insight-based evaluation model is far more applicable than traditional benchmark metrics such as task error rates and task completion time. 
The evaluation framework by Fragiadakis et al.~\cite{fragiadakis2025evaluating} supports a structured analysis of Human-AI collaboration (AI-centric, Human-centric, and symbiotic) by identifying relevant evaluation metrics, but does not focus on multimedia nor visual analytics aspects.
Finally, the AQ model by Zahálka et al.~\cite{zahalka2015analytic} models human thinking as highly flexible \emph{analytic categorization}, with the human dynamically adding, removing, or updating/redefining categories~\cite{zahalka2014towards} with measures for user insight gain and acquisition time. Categorization is a useful abstraction covering several tasks. Full evaluation of multimedia analytics remains an open problem. 

\section{Framework Design}
\label{sec:framework_construction}

We now derive our design principles which are structured to ensure conceptual coherence (\Dconsist), technical realization and robustness (\Dfunction~\Dlimit~\Dvisualize), and a combined human-AI perspective (\Dinteract \Dshift). 

Models for visual / multimedia analytics and human computer interaction as described in related work have been developed for decades.  Many of them such as the canonical separation into data-models-visualization-knowledge, the iterative sensemaking of the knowledge generation model, Munzner's task abstractions, and the guidance model are to a large extent still relevant in the foundation model era. The shift to agentic AI that FMs initiated requires integrating, re-grounding, and adapting them to the agentic era:
\begin{itemize}[align=left, itemsep=0pt, parsep=0pt]
    \item[\Dconsist] {\ttfamily Consistency and Integration:} Align with well established frameworks for Visual Analytics, Multimedia Analytics, and HCI -- building upon them as well as integrating them whenever appropriate -- but adapt to capture the more agentic role of AI.
\end{itemize}

The advancements in AI models are extremely rapid. A successful framework must fully embrace these opportunities and ensure optimized instructions to maximize the output quality of AI agents. To allow drop-in replacement of new FMs and AI agents as they emerge, a modular agentic design with clear task decomposition and abstraction mechanisms to access the modules, as well as clearly defined orchestration are essential:

\begin{itemize}[align=left, itemsep=0pt, parsep=0pt]
    \item[\Dfunction] {\ttfamily AI Functionality:} Be able to easily adapt to current and future AI functionalities in terms of multimodal analysis, reasoning abilities, and autonomous behavior.
\end{itemize}

Despite the successes of FMs, AI still has several limitations, most prominently hallucinations, bias, lack of human values, tunnel vision, and lack of transparency. Some of these can be solved by further improvements of FMs or agents using them, but many of them can inherently only be solved by bringing in human expertise. To make this possible, the framework should provide the user with ample opportunities to steer the session directly. On the other hand, the AI should provide the user with a rationale explaining which reasoning steps were used to get to the answer and which source was used and why, guiding the user towards verifying it directly:

\begin{itemize}[align=left, itemsep=0pt, parsep=0pt]
    \item[\Dlimit] {\ttfamily AI Limitations:} Address the inherent limitations of AI due to their architecture, training data, and procedure, through human-AI teaming based solutions.
\end{itemize}

Traditional visual analytics solutions assume that the human expert is in control of the analytic process and has access to a model with clear characteristics and model parameters that can be tuned. The shift to AI based models leaves many design criteria like balancing conciseness, completeness, and being understandable intact. However, how to operationalize them for a specific model requires substantial rethinking. Due to the AI limitations described above, the truthfulness of the results is not evident anymore and systems have to make the trustworthiness tangible. Finally, the analytical workflow has become more complex, and tasks might be executed in parallel in an autonomous way with a partial shift for the user from acting to monitoring. This leads to:

\begin{itemize}[align=left, itemsep=0pt, parsep=0pt]
    \item[\Dvisualize] {\ttfamily Visualization:} The expert user should be provided with all information relevant to the task in a concise, complete, understandable, and truthful manner and receive guidance to optimize the parallel analytical workflow. 
\end{itemize}
\label{sec:development}

Following established VA theory~\cite{pike2009science,zahalka2014towards}, interactions should be intrinsic to the framework, in particular sequences of interaction loops should translate high-level user intent into atomic analytic tasks to be processed by the system where different actions are clearly separated in terms of data and knowledge used and actions operating on them. Timing should be addressed explicitly using an orchestration fitting the use of agents so users are not blocked in their analytic process. To reinforce user trust and patience, transparency of the AI explaining what is being processed and how long this will take is important. Finally, communication between the human and heterogeneous AI agents and tools should be expressive while at the same time structured to remain comprehensible for both human and AI. In summary:

\begin{itemize}[align=left, itemsep=0pt, parsep=0pt]
    \item[\Dinteract] {\ttfamily Interaction Channels:} Exhibit a seamless, explicitly separable, interaction channel between users and AI that considers data and knowledge flow, technical implementation patterns, and timeliness.
\end{itemize}

Many analytic tasks for which human intelligence was the only solution can now be done well and in various cases better by AI. In order to enable this shift in analytical reasoning, our framework should leverage the potential collaborative nature of agentic AI. Mechanisms are needed to externalize user reasoning, enable steering, autonomy delegation, and hypothesis generation. These mechanisms should provide explicit records of decision provenance, shifting reasoning from human-driven to true shared human-AI collaboration that is both transparent and auditable. This also requires to make attribution explicit by distinguishing different sources of data and knowledge, allocating them across humans, FMs and expert modules for provenance, privacy, and bias evaluation. Finally, the human-AI teaming principles i.e. mutual goals, co-management, and progress tracking should have  concrete UI hooks. The above leads to:

\begin{itemize}[align=left, itemsep=0pt, parsep=0pt]
    \item[\Dshift] {\ttfamily Shift in Analytical Reasoning:}
    Capture the collaborative nature of agentic AI and effects on steering, autonomy, knowledge generation, attribution, privacy, and provenance, both in processing and user interfaces.
\end{itemize}

The above criteria are only partially met in sufficient detail by existing VA / MMA frameworks, most of which are predating the rise of foundation models and agentic AI. Consequently, many models implicitly assume a user in ultimate command, steering the exploration process and controlling narrowly-defined, often silo-ed algorithmic modules, thereby lacking a process-centric, technical view and having difficulties in clearly describing the emerging role of FMs acting as \emph{true collaborative partners}~\cite{Hutchinson.FoundationalVAOpportunities.2025}.


Our framework design process was conducted over the course of nearly hundred person-hours of discussions and iterations among the authors, a mix of established researchers in multimedia and visual analytics. With the design criteria as starting point, drawings of the architecture and its components were created by considering established models and reviews, identifying the main building blocks, processes, attribution of data and knowledge, and communication channels. In an iterative manner, we then added, refined, and discarded aspects, before converging on an agreed and consistent version.

\hide{
\begin{table*}[ht]
\centering

\caption{Analytic limitations of Foundation Models, and their mitigations (\autoref{sec:hybrid_mitigation}). }
\label{tab:fm_limitations}
\fontsize{8}{10}\selectfont
\renewcommand{\arraystretch}{1.1}
\begin{tabularx}{\textwidth}{@{}p{3.1cm}@{\hskip3pt}p{10.4cm}@{\hskip6pt}p{4.4cm}@{}}
\toprule
\textbf{Limitation} & \textbf{System mitigation} & \textbf{Human guidance} \\
\midrule
\textbf{Insufficient knowledge} & Retrieval augmented generation (RAG)~\cite{Patrick2020RAGforKnowledgeIntensive, zhao2023-retrieving}, expert modules~\cite{karpas2022mrklsystemsmodularneurosymbolic} & Knowledge injection by interaction\\
\textbf{Hallucinations} & Strict RAG rules + citation enforcement~\cite{Huang2025hallucination}, citation checking~\cite{cohen2024contextcite}, uncertainty detection~\cite{Huang2025hallucination} & Information source verification \\
\textbf{Analytic tasks complexity} & Task decomposition using agentic workflow~\cite{mcp, Guo2024MultiAgentSurvey890,Xi2025LLMagentsurvey}, reasoning models~\cite{kojima2022large,wei2022chain} & Explicit decomposition and steering \\
\textbf{Tunnel vision} & Reasoning models~\cite{kojima2022large,wei2022chain}, domain-specific instructions~\cite{Patrick2020RAGforKnowledgeIntensive, zhao2023-retrieving}, multi-agent teams~\cite{Guo2024MultiAgentSurvey890,Sperrle.CoAdaptiveGuidance.2021} & Reasoning and focus guidance\\
\textbf{Limited comm. channels} & Agentic workflow~\cite{mcp, Guo2024MultiAgentSurvey890,Xi2025LLMagentsurvey} directly manipulating visual analytics environments & Rich interactions\\
\textbf{Bias/lack of human values} & Guardrails~\cite{inan2023llama,schick2021self,zhao2021calibrate}, expert modules~\cite{karpas2022mrklsystemsmodularneurosymbolic}, agentic workflow~\cite{mcp, Guo2024MultiAgentSurvey890,Xi2025LLMagentsurvey} & Human oversight, evaluation audits \\
\textbf{Provider dependency} & Local models, Curated datasets & Increased human involvement \\
\bottomrule
\end{tabularx}
\vspace*{-2mm}
\end{table*}
}

\section{Foundation Model-based AI \Dfunction\Dlimit\Dshift}

\label{sec:AImodels}

\subsection{Knowledge and Data \Dshift}

Before deep learning, there was always a clear separation between the unstructured data sources and the \textbf{factual knowledge} captured in explicit representations like knowledge graphs. Foundation models have \textbf{implicit knowledge} represented in local patterns of model parameters. 
For the data, there is another important distinguishing factor. The FM has been trained on \textbf{generic data} captured in large volumes,  while in many applications there are also a lot of \textbf{domain specific data} that the FM has not seen due to non-public availability or because the public data came available after the last training of the model. To access factual knowledge and domain specific data, additional mechanisms are needed. 

\subsection{AI Agents \Dfunction}
\label{sec:ai_agents}

An FM can be used directly, but also be a core component of an AI agent capable of autonomously performing a specific task for users~\cite{Guo2024MultiAgentSurvey890,Xi2025LLMagentsurvey}. The agent can perform its given task using explicit reasoning and planning abilities, a clear chain of thought, and the decomposition of the problem into subproblems. There are many different agents~\cite{casper2025aiagentindex} and many ways to structure the set of possible agents. We follow the distinctions made by Schulhoff et al.~\cite{schulhoff2025promptreportsystematicsurvey}.
\parheader{Code Generation Agents} 
Rather than answering queries, FMs can now also be instructed to provide a piece of software code to answer or perform a specific task. This can be the target users are pursuing, but it can also provide code to create visualizations and even complete dashboards.    
\parheader{Observation-Based Agents}
Agents can monitor sensors, websites, or processes, and take actions based on that. The observations become part of the prompt and the agent can be instructed to only generate output when specific conditions are met.
\parheader{Retrieval Augmented Generation (RAG)}
To inject domain knowledge into the analytics process RAG is widely used~\cite{Patrick2020RAGforKnowledgeIntensive}. RAG allows for keyword and/or semantic search of relevant context from \textbf{domain-specific data} and \textbf{factual knowledge} necessary for generating correct responses. Zhao et al. survey how additional modalities can improve upon a text-only RAG approach~\cite{zhao2023-retrieving}. Yasunga et al. present a more symmetric approach where they consider multimodal RAGs for both the caption-to-image and image-to-caption tasks~\cite{Yasunga2023RAGMultimodal}. The trustworthiness and quality of external knowledge still require attention~\cite{Fa2024RAGmeetingLLM}.

Being able to use external tools is one of the defining features of an AI agent: the AI goes from merely producing outputs to truly interacting with the world. Efficient communication, including connections to tools and other software, can be supported using the Model Context Protocol (MCP)~\cite{mcp}. Especially important are \textbf{expert modules}~\cite{karpas2022mrklsystemsmodularneurosymbolic} which have specific functionality and explicit factual knowledge ranging from e.g. a calculator performing basic operations, to advanced systems using a domain-specific knowledge graph. Expert modules are curated by human domain experts, and the information is adequately sourced. This provides a solid basis for high-quality agent answers. Consequently, it's hard to create expert modules, and the volume of information they can provide is much lower than the FM training data. Agentic systems should marry the quality of expert module responses with the broadness of the foundation model.

\subsection{Limitations of Foundation Models \Dlimit}
\label{sec:fm_limitations}

There are still several important limitations that preclude direct deployment of FMs to independently solve high-stakes analytic tasks: 
    \parheader{Insufficient knowledge} FMs are trained on massive, web-scale datasets, and therefore possess a large body of \emph{implicit knowledge}. They lack recent information, are trained to be generic, lacking  highly specialized \emph{factual knowledge}, and proprietary information. 
    \parheader{Hallucinations} Generative AI models can present plausible, yet nonfactual, content to users as an answer, in particular discrepancies between the generated content and verifiable real-world facts, divergence of generated content from user input, or a lack of self-consistency within the generated content~\cite{Huang2025hallucination}.
    \parheader{Complexity of analytic tasks} Many analytic tasks are complex, requiring a large number of steps. FMs are eager to provide a solution right away, which is counterproductive, as they may cut corners. As more context is added, they become lost in the provided information and consequently lose their problem-solving skills. The model may also suffer from context degradation: ignoring specific prompt instructions and reverting to its vanilla behavior~\cite{liu2024lost,zhao2021calibrate}.
    \parheader{Tunnel vision} FMs have a tendency to only consider a single strategy at a given time~\cite{yao2023tree} and sticking to it to the last possible moment,  and then abandon it completely at the slightest hint of skepticism from the user. This stems from the model's fundamental nature as attention works best when focusing on a single narrative~\cite{vaswani2017attention}. 
    \parheader{Limited communication channel} Despite the recent advances in agentic protocols such as MCP~\cite{mcp}, the textual channel (possibly with domain-specific context) remains the main communication channel. Modern agentic AI can create a wealth of useful outputs (including charts, documents, or spreadsheets) but agents have no inherent knowledge of VA so active iterative conversations, typical for analytics, still rely on the textual channel.
    \parheader{Bias and lack of human values} FM outputs may introduce biases or inaccuracies, emphasizing the need for human validation~\cite{fabbrizzi2022bias,Tang2022VideoModerator,FiHiJe2022VAIntelligence}. Furthermore, an AI model often does not adhere to the RICE characteristics~\cite{ji2024aialignmentcomprehensivesurvey}: \emph{robustness} (operates reliably under diverse scenarios and is resilient to unforeseen disruptions);  \emph{interpretability} (decisions and intentions are comprehensible and reasoning is unconcealed and truthful); \emph{controllability} (behaviors can be directed by humans and allows human intervention when needed); and \emph{ethicality} (adheres to global moral standards and respects values within human society).
    \parheader{Provider dependency} Full training of FMs can only be done by the major tech companies. Consequently, such models can often only be accessed using cloud-based solutions and organizations using them have no control over the data on which those models are trained. For many (government) organizations such solutions are not an option.

\subsection{Prompt Templates \Dfunction}
\label{sec:prompt-templates}
\hide{\todo{Emphasize how prompt templates help in adapting to improved functionalities.}}
The input to the AI models is a \emph{prompt}, which is primarily based on text but can contain images, sound, or other media. In principle, a prompt is in a free format, but structure can be added. \emph{Directives} are a textual specification of the user's intent that can be elaborated on by giving a number of \emph{examples}. Specifying \emph{output formatting} with optional \emph{style instructions} makes the type of response explicit. To personalize output, an explicit \emph{role} can be specified. Finally, the system might require the user to give \emph{additional information} to give the proper output. 

The free format prompts give users the freedom to express the same thing using slightly different prompts or have a similar prompt with a different meaning. This makes it difficult to reason on the AI models at a more abstract level, e.g., doing prompt engineering to create better prompts, making an informed choice for a specific agent, or defining a strategy containing a number of subsequent prompts to reach some desired target. 
On the AI side, we see a similar phenomenon. Having a mix of AI agents, foundation models, and external tools, there are various ways to provide specific functionality to the user. For the user, the quality of the response and a trustworthy rationale for how the result is obtained, are far more important than how the tools are internally organized. To add robustness, we follow the comprehensive survey by Schulhoff et al.~\cite{schulhoff2025promptreportsystematicsurvey} which considers the use of prompt templates as an abstraction mechanism to create various prompts. Extending the definition in the survey~\cite{schulhoff2025promptreportsystematicsurvey}, by explicitly making the abstraction a contract, we define:

\conceptdefinition{Prompt template}{a function containing one or more variables to be replaced by some media (text, image, video, sound, or other) to create a \emph{structured prompt} as an instance of the template and thus defining a contract between the action specific agent making the \emph{(multimodal) request} and the collective functionality of the AI models responding as one virtual model with \emph{structured output} and a \emph{rationale}.}

\noindent The existing set of templates~\cite{schulhoff2025promptreportsystematicsurvey} is based on 33 prompt vocabulary terms, a taxonomy of 58 language-based techniques, and 40 techniques for other modalities.  Clearly, the set of agents~\cite{casper2025aiagentindex}, FMs and external modules will evolve continuously with best task performance shifting from one type of AI tool to the other. The corresponding set of functionality based prompt templates will grow dynamically to adapt, albeit at a much lower pace of change. 

\section{Human Understanding}
\label{sec:human_analytics}

\subsection{Knowledge Generation \Dconsist\Dshift}
\label{sec:knowledge_generation}

In analytics, users typically start with \emph{hypotheses} grounded in their prior \emph{knowledge} about the domain. This prior knowledge not only frames what is considered plausible, but also guides the selection of relevant data, methods, and representations. Expert users aim to address these hypotheses through executing and accomplishing high-level \emph{tasks}. These tasks consist of a sequence of \emph{actions} that support analytic reasoning and insight generation, where much of the latter occurs as part of human cognition, which has been conceptualized in the Knowledge Generation Model~\cite{sacha2014knowlegde}. The sequence of actions is based on user intent~\cite{Yi2007interaction} and can be translated from domain language into more abstract structures~\cite{munzner2015visualization} and combined with user-intent based interaction techniques~\cite{Yi2007interaction}.
The execution of the user tasks leads to \emph{findings} and \emph{insights}: findings are local observations tied to particular data and views, while insights represent more coherent and generalized conclusions, which contribute to more durable \emph{knowledge}.  Crucially, in the Knowledge Generation Model, knowledge is not merely the end product but also a continuous resource: it shapes new hypotheses, influences subsequent exploration, and is externalized through annotations, reports, or encoded models so that it can be communicated, reused, and built upon. VA actions are categorized as \emph{Analysis, Search, and Query}~\cite{munzner2015visualization} which for FMs need to be extended with \emph{Generation}~\cite{Lin2024evaluatingGenAI}. Structuring a collection through various categorizations can be viewed as a rudimentary form of knowledge \cite{zahalka2014towards}. Subsequent knowledge generation can then focus on adding (hyper)graph structures on top of the data and the categories \cite{gisolf2025hypergraphs}. The model in~\cite{sacha2014knowlegde} assumes an expert user who is in full control, but the AI model has access to both factual and implicit knowledge and can reason to create new knowledge. A knowledge base shared by AI models and the expert user is needed in which knowledge and its provenance are made explicit.

\subsection{User Interface \Dvisualize\Dinteract\Dshift}
\label{sec:ui}

\begin{figure}[h]
  \centering
  \includegraphics[width=1.1\linewidth]{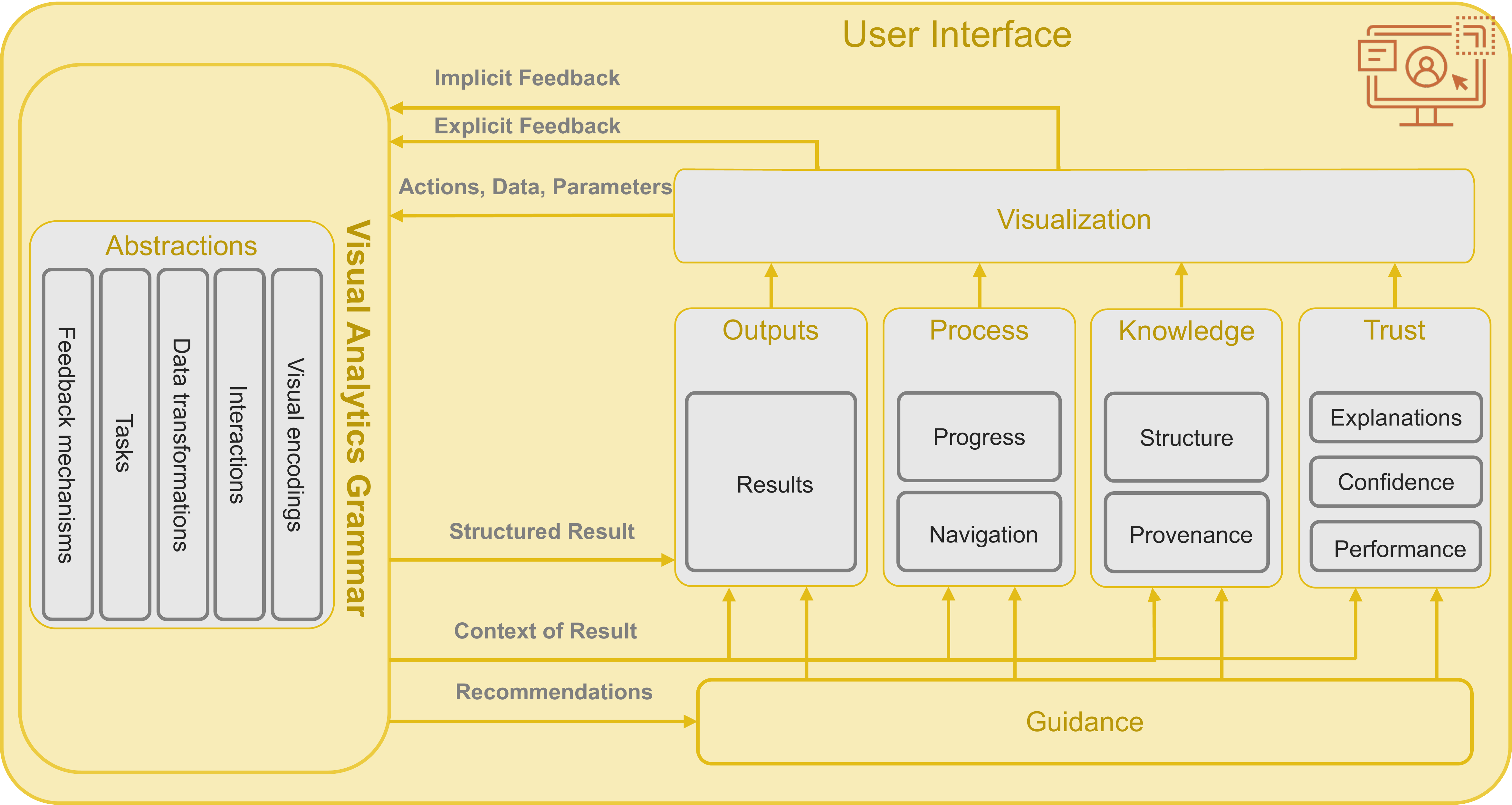}
  \caption{%
    \bgcolorbox[ColorHumanAITeaming!30]{User Interface}, model zoomed-in from \autoref{fig:teaser}, showing in an abstract manner the main components a user interface should have in the model. The visual analytics grammar forms the communication channel with the human-AI teaming.
  \label{fig:user-interface}
  }
  \vspace*{-2mm}
\end{figure}

Many current human-AI interfaces focus on textual inputs, often directly inputting prompts. In line with Chen et al.~\cite{chen2025interchat} we argue that the UI for foundation models should  contain a strong visualization component. Any specific UI would be application dependent. Here, we therefore focus on the generic components an interface should have and how these components connect to the other elements (see~\autoref{fig:user-interface}). An interface conforming to the framework has four pillars; \emph{outputs, process, knowledge}, and \emph{trust} and we argue that these should  be supported with rich interactive visualizations also providing the handles to give implicit and explicit feedback to the AI models which can then give guidance to the users. 

\textbf{Outputs} are the main result of the VA agents and come with directives for visualizing them. Naturally, users are enabled to change idioms through interaction for additional perspective exploration. Guidance can, for example, highlight interesting patterns in the results, or indicate areas of the information space containing underexplored data. The \textbf{process}
component provides means to \emph{navigate} large-scale multimedia datasets more efficiently, while simultaneously showing the \emph{progress} towards the goals (e.g., by progressive visualization~\cite{Ulmer2024Progressive}), particularly important because of the asynchronous nature of AI agents. Guidance can be in the form of suggesting the next step or suggesting strategies to reach the goal more efficiently. \textbf{Knowledge} should be visualized with specific attention for the provenance of knowledge; was it the human or AI who provided the knowledge and was it implicit or factual? To give guidance, the VA agents can suggest new or delete existing relationships that the user can confirm. Finally, to gain \textbf{trust} the user should receive clear explanations of the rationale used by the AI model, reliable measures of confidence in the results, and depictions of performance. 

Similarly to how prompt templates encode structured ways to work with AI models, a \emph{visual analytics grammar} defines abstract representations of the interface components and their interactions, e.g. a multimedia extension of~\cite{Satyanarayan2017vega} and \cite{AndrienkoATWL2025}, explicitly taking into account that for assessment of images and videos they need to be clearly visible to the user. The grammar hides implementation details, while providing sufficient information to reason about how to best present the results in a structured way, the right handles for the user to give feedback, and for the AI to give recommendations. The grammar allows to express valid workflows, where different low-level implementations can be used:

\conceptdefinition{Visual analytics grammar}{An explicit broad enough, yet machine-readable, description defining high level valid analytic workflows in visual analytic systems through API-like machine interfaces of visual encodings, interaction techniques, data transformations, tasks, and feedback mechanisms.}

\begin{figure}[b]
  \centering
  \includegraphics[width=1.0\linewidth]{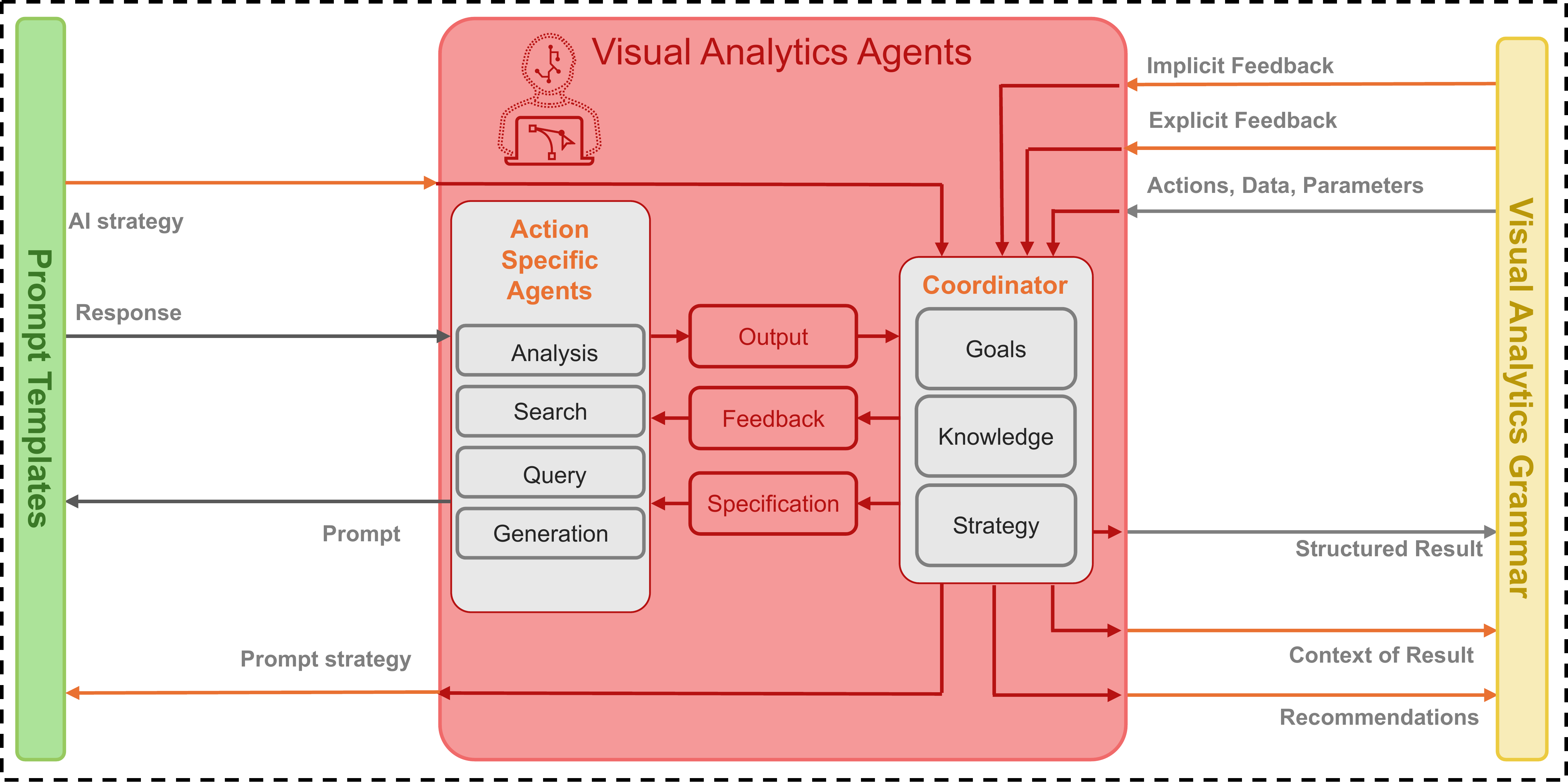}
  \caption{%
    \bgcolorbox[ColorHumanAITeaming!30]{Human-AI Teaming} model, zoomed-in from \autoref{fig:teaser}, conceptualizing how visual analytics agents let users communicate in an effective way via a targeted user interface with the FMs and \textcolor{ColorAIAgent!40!black}{AI agents}.}
  \label{fig:human-in-the-loop}
  \vspace*{-2mm}
\end{figure}

\section{Human-AI Teaming \Dlimit\Dinteract\Dshift}
\label{sec:human-ai-teaming}

To rigorously structure human-AI teaming (see \autoref{fig:human-in-the-loop}) and to allow for the shift in analytic reasoning from humans to AI, we define specialized VA agents communicating with AI models through prompt templates and with users via the visual analytics grammar.

\subsection{Visual Analytics Agents~\Dconsist~\Dshift}
\label{sec:va_agents}

\begin{figure}[h]
\centering
    \includegraphics[width=\linewidth]{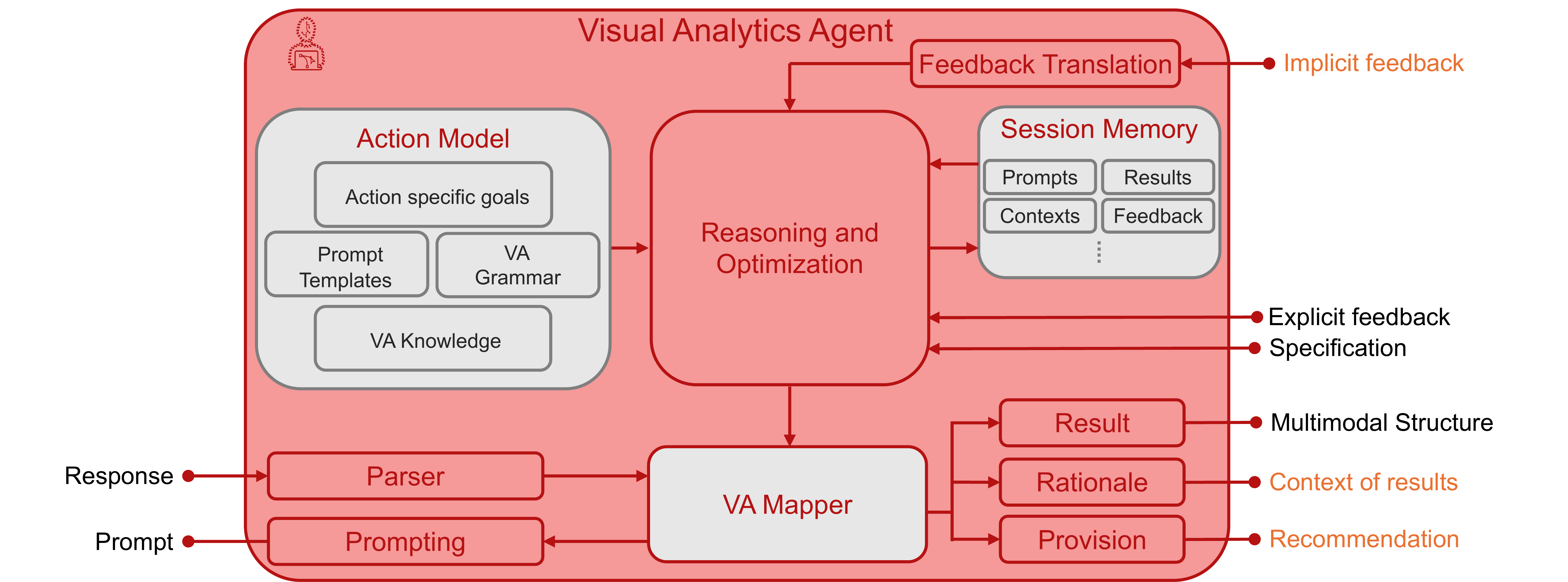}
  \caption{%
    Schematic representation of the generic architecture of a the visual analytics agents in \autoref{fig:human-in-the-loop}.  
    }
  \label{fig:va-agent}
  \vspace*{-2mm}
\end{figure}

Visual analytics agents (see \autoref{fig:va-agent}) share many of the characteristics of the AI agents we considered earlier with some specific characteristics. They are specialized for a specific type of action as identified in \autoref{sec:human_analytics} i.e. analysis, search, query, and generation and have knowledge of visual analytics considering both the processing of user requests and the best way to present the results, explicitly optimizing the mapping between humans and AI in a single interaction cycle and at the session level. 

The starting point for the definition of VA agents is the \textbf{action model} with an explicitly defined action goal and a quantitative measure to assess progress towards the goal. The model has an action specific set of visual analytics grammar expressions and prompt templates (see \autoref{sec:ui} and \autoref{sec:prompt-templates}). The  \textbf{VA knowledge} comprises elements such as capabilities and limitations of human cognition, visualization design rules, trust building methods, and interactive learning strategies. 

The main human-AI interaction channel is the \textbf{VA mapper} which provides three mappings. The $<$visual analytics grammar $\rightarrow$ prompt template$>$ mapping takes a specification by the user via the interface and explicit feedback on the previous iterations and selects the right prompt templates. The results from the AI are parsed and aggregated to suit further processing. The $<$response $\rightarrow$ visual analytics grammar$>$ mapping, steered by the VA knowledge, selects the best way to present the (multimodal) result and its context. The final mapping is the $<$feedback $\rightarrow$ provision$>$ mapping, which based on the results obtained up to this point and the explicit and implicit user feedback (which needs translation~\cite{perez2022typology}), provides recommendations to the user to reach the goal effectively. For finding the optimal mappings, there are different ways to proceed. From the perspective of data and results obtained, the main techniques are active learning, interactive machine learning, and machine teaching~\cite{jiang2019recent,mosqueira2023human}.  Alternatively, the VA agent can actively try different prompts, and optimize them according to the results~\cite{schulhoff2025promptreportsystematicsurvey}. At session level, such techniques are typically combined with an explicit strategy and a \textbf{session memory} of all previous steps optimized for efficient retrieval of relevant context. This sets the stage for using human-in-the-loop reinforcement learning~\cite{retzlaff2024human}.

Having multiple agents requires a control architecture which could be hierarchical, centralized, decentralized, or having a shared message pool \cite{Sheng2026MultiAgent}. In our framework, we consider the human as being in ultimate control and hence the central coordinator. We therefore add a \textbf{coordinator agent} abstracting the user as an agent controlling the  whole process. This agent has the same architecture as the other VA agents, but operates on a different level. The coordinator will have an overarching goal which it has to decompose into subgoals, or let the user define them as such, to be addressed by one of the action-specific agents, the results of which are aggregated. The VA strategy is geared toward reaching the overarching goal, hence operating on a more abstract level. The VA agents work in an asynchronous manner and thus the coordinator should ensure independence of the individual parallel actions or order them, and should take into account that results might come back at different times. 

\subsection{Strategy loop \Dconsist~\Dinteract~\Dshift}
\label{sec:strategy_loop}

As indicated in \autoref{fig:teaser}, the FM-based AI connects to the  human-AI teaming through the \emph{strategy loop}.  Through this high level loop the AI agents carry out analytic strategies delegated by the human-AI teaming and in addition to the results should generate explanations for their internal multi-step strategies. There are many different strategies, for prompt based interaction often referred to as chain of thought (CoT), that could be considered, where \cite{Maciej_graph_of_thought2024} distinguishes single CoT, multiple CoT, tree of thoughts, and graph of thoughts. The prompting strategy to use depends on the complexity of the task and has a direct effect on the communication between the human-AI teaming and the AI models, but also on how these strategies are communicated visually to the user.  

At the single interaction level, the \emph{responding-prompting loop} operationalizes the above communication with the human-AI teaming querying with structured prompts and AI models producing rationalized responses for each step within the broader context of the strategy.
The optimization of the workflow operates on two levels, namely, at the individual action-based agent level and at the coordinating agent level. In a reinforcement learning environment~\cite{Ma_reinforcement_retrieval_2024,metz2025mappingspacehumanfeedback,retzlaff2024human}, the core is to define and optimize the respective reward models. As for the agent observing the human by interactions might not be enough, this requires interface components to define explicit goals that can be shared with the agents. The decomposition into different actions helps here, but a balance has to be found. Users should not be bothered with too many fixed goals, as they might change during the course of the analytic process, whereas for the system explicit goals are helpful in determining an optimal strategy. The workflow should be supported by a set of interactive visualizations, abstracted in the visual analytics grammar, to help the user perform all tasks while providing the system with the feedback needed for optimization.

\subsection{Guidance and Trust loop \Dinteract}
\label{sec:trust-building-loop}

The low-level interaction between users and VA agents is the \emph{specification-result loop} where users formulate their query and the relevant \textbf{actions, data, and parameters}. The \textbf{structured result} is a multimodal structure expressed in visual analytics grammar elements. The quality of the loop is determined by the accuracy and completeness of the answers. To steer the agents, users give \textbf{explicit feedback} on the results. The \textbf{coordinator agent} component forwards this feedback to the appropriate VA agent to enable interactive learning, procedural guidance, personalization, and harnessing the user's intuition.

To increase transparency and build trust, the UI puts the rationale of the VA agents into \textbf{context} (chain-of-thought, uncertainty, performance), so users can identify potential issues or verify the output before taking actions.  
Explainability for FMs is challenging, but several techniques can help. Attention visualization (e.g.,~\cite{Li2023Attention}), shows which parts of an image, video, or text influenced the AI output. Counterfactual explanations indicate what changes would have led to a different output. Explicitly decomposing complex AI agent tasks enables explanation and oversight of intermediate steps. The interface should also show uncertainty, performance, and confidence as context for the output. As our framework has specific VA agents for the four actions, each providing a step-by-step rationale, it is easier for the user to grasp the overall rationale.

To solve analytic problems through human-AI teaming, the system should provision \textbf{recommendations} to address the user's “knowledge gaps” by identifying them and providing orienting, directing, and prescriptive guidance~\cite{perez2022typology}. These three types of guidance differ in freedom. In orienting guidance, the user is given some recommendations as support. Prescriptive guidance is the other extreme, where the user can only accept or disregard the disruptive guidance completely. Directing guidance provides several options to choose from. The system should also gather \textbf{implicit feedback}  on what guidance to provide by observation of the user-system interactions, storing and translating this feedback into a form that can be used for strategy optimization. In~\cite{metz2025mappingspacehumanfeedback} a categorization of the types of human feedback given. It could be \emph{evaluative}, giving a direct assessment of the performance of the agent, \emph{instructive} where the user gives explicit hints to the agent on how to perform the action, or \emph{descriptive} giving additional information to the agent that can help perform the action.  

\subsection{Hybrid Mitigation of FM Limitations~\Dlimit}
\label{sec:hybrid_mitigation}

We now iterate on the AI limitations~(\autoref{sec:fm_limitations}) and discuss how human-AI teaming can combine existing system mitigations on the machine side and human guidance aspects.

\parheader{Insufficient knowledge} 
RAG~\cite{Patrick2020RAGforKnowledgeIntensive, zhao2023-retrieving} augments incoming prompts with context retrieved from a database. When the agent cannot retrieve the necessary context, it should consult the user, where at the same time, the user can ask the agent to base its reasoning on a specific assumption or piece of knowledge.
\parheader{Hallucinations}
We can condition RAG to strictly use the knowledge base when answering domain-specific questions~\cite{Huang2025hallucination}, using external citation checking or uncertainty checking~\cite{cohen2024contextcite}\cite{Huang2025hallucination} or a verifier agent~\cite{Gou2024CRITIC}. The user plays an essential role in verifying the trustworthiness of the external sources deployed, restricting external sources to a pre-defined set, or validating cases where the verification result is ambiguous.  
\parheader{Complexity of analytic tasks} 
Analytic tasks are often complex, precluding a direct, single-prompt solution and might be tackled in the system by agentic workflows~\cite{Guo2024MultiAgentSurvey890,Xi2025LLMagentsurvey}.The agent and the user should jointly decompose complex tasks or parts of it, reflect on the intermediate steps, and be able to steer the overall decomposition. 
\parheader{Tunnel vision} 
There are three techniques to alleviate tunnel vision on the model side: domain-specific instructions, prompt engineering in RAG~\cite{Patrick2020RAGforKnowledgeIntensive, zhao2023-retrieving}, and leveraging multi-agent teams. Reasoning models~\cite{kojima2022large,wei2022chain} provide transparency and invite user intervention wherever needed. Timely user interventions prevent long chains of thought in unproductive directions. 
\parheader{Limited communication channels}
Using text as the communication channel is often chosen as it is the most basic channel familiar to both naive and expert users. The Model Context Protocol (MCP)~\cite{mcp} provides the necessary flexibility to address user requests in diverse analytics scenarios. This has to be complemented at the user side with a visual analytics grammar to allow much richer interaction. 
\parheader{Bias and lack of human values}
is intrinsically tied to model training and/or finetuning~\cite{ouyang2022training,stiennon2020learning,wang2022exploring}. Techniques exist, however, that allow to mitigate bias post-training~\cite{schick2021self}\cite{zhao2021calibrate}\cite{inan2023llama}. In sensitive domains, the user should always provide oversight and strong feedback taking into account the context and assess bias and ethical considerations accordingly. 
\parheader{Provider dependency} When using simpler local models, the balance between human and AI shifts more to the expert user, making the role of visual analytics even more important. 

\section{Evaluation of Analytics Solutions}

\label{sec:evaluation}

Below, we discuss for each main component of our model, which evaluation metrics could be applied or should be developed.

\parheader{Foundation model-based AI} Benchmarks for the analytic capabilities of FMs are very limited. 
\emph{AI agents} build primarily on reinforcement learning (RL), and its evaluation typically focuses on an agent’s ability to complete tasks within a well defined environment~\cite{ji2023safety,lin2023mcu,yuan2023rl}. While these efforts advance evaluation in specialized settings, there is a need for analytics-tailored frameworks to support systematic and transferable assessment across tasks and domains.
\parheader{Human understanding} Quantitative methods to evaluate \emph{User interfaces} help to identify efficiency (e.g., task completion time), effectiveness (e.g., error rate), and cognitive load (e.g, eye-tracking, NASA-TLX~\cite{Hart1988NasaTLX}).  To evaluate the cognitive aspects, insight based methods are far more appropriate than benchmarks~\cite{north2006toward}. A tailored  evaluation model incorporating multimedia, based on a visual analytics grammar, is missing.
\parheader{Human-AI teaming} Evaluating the human-AI teaming core of analytics systems remains an open challenge as this will need evaluation methods that take into account (a chain of) multiple components rather than an isolated task that can easily be benchmarked. Insight-based methods~\cite{north2006toward} are useful, but at the same time we want to be able to automatically evaluate VA agents in terms of performance, speed, scalability, robustness, explainability, and fairness. Analytic Quality (AQ)~\cite{zahalka2015analytic} seeks to bridge this gap by simulating user evaluation runs of multimedia analytics backends and measuring performance and efficiency metrics. This approach, however, is not directly applicable in the foundation model era, with new paradigms of interaction and reasoning. The evaluation of conversational visual analytics where data analysis is enabled through natural language \cite{Palani2026Lexara} might be an inspiration here. For evaluation the video browser showdown \cite{gia2025vbs,Lokuc2023VBS} is highly appropriate, but has a focus on retrieval rather than analytics. New evaluation paradigms are needed.  

\section{Impact and conclusion}
\label{sec:conclusion}
Our framework combines the strengths of modern AI for multimedia analysis and visual analytics to anticipate a future in which human and AI form teams to reach ambitious goals. The definition sets the scene for innovative research ranging from developing a visual analytics grammar abstracting user interactions and visualization, visual analytics agents that optimize workflows and build trust, and new evaluation methods for human-AI teaming. The framework suggests a new way of designing multimedia analytics applications: 
1)~Specification  of the high level goal(s) of the systems and decomposition of the goal(s) into the four actions (\autoref{sec:knowledge_generation});
2)~Selection of the set of relevant prompt templates (\autoref{sec:prompt-templates});
3)~Expressing an abstract user interface in the visual analytics grammar (\autoref{sec:ui}); 
4)~Inventory of implicit and explicit feedback from the user VA agents can use and the recommendations to provide to the user (\autoref{sec:trust-building-loop});
5)~Definition of strategies and optimization (\autoref{sec:strategy_loop}).
We will now consider how the framework might impact some typical multimedia applications. 


Recent work in making cultural heritage accessible has focused on LLM based methods to obtain comprehensive descriptions of paintings \cite{Bin_GalleryGPT_2025,Rachabatuni_artChatBot_2024,Shuai_artRag_2025}. Advanced visualizations would enable going beyond individual paintings and allow art historians to study trends, influences, and relations among artists at scale. Graph- and hypergraph-based visualizations are already prominent, in \cite{Fischer.MultiCase.2024} for multimodal evidence in police investigations and in   \cite{gisolf2025hypergraphs} focusing on images in investigating incidents like MH17, using pre-computed information like visual concepts and named entities. The framework would suggest a shift to dynamic methods which allow for on-the-fly information extraction and incremental understanding, with a balance of human and AI driven analytics. The current performance of LLM based methods for challenging multimodal financial data \cite{Zhu_multimodal_finance2025} also suggests that for a long time to come, human-AI teaming will be essential. To that end, interactive visualizations to aid in understanding and building trust in  the AI tools are necessary. Access to large video archives from a user perspective is actively pursued in the Video Browser Showdown \cite{Lokuc2023VBS} and in the Castle challenge \cite{Khan_Castle_2025}. While the focus of these challenges is on retrieval with methods focusing on objects \cite{Prak_VBS2026}, temporal queries and relevance feedback \cite{Exquisitor_VBS2026}, or query expansion using text and image generation~\cite{NII-UIT-VBS2026}, recent editions have started to focus on Video Question answering (VQA) which requires basic analytic reasoning. In \cite{Meng_GraphVideoAgent_2025} VQA  for long videos is considered. In current retrieval targeted systems, the interfaces are predominantly based on image grids although for the rich Castle dataset more elaborate visualizations are being explored \cite{Khan_Castle_2025}. Future incarnations of the challenges could consider far more complicated and realistic questions which require accessing much larger parts of the dataset and complex relations among the video fragments. The proposed framework would then allow leveraging the already strong retrieval components and employing them in a true multimedia analytics setting. A system explicitly targeting the generation of images is PromptMagician \cite{Feng2024promptmagician} which aids users in creating images fitting their preferences. This kind of system is an interesting component of future systems in which generation is one of the components in the overall system working alongside agents for analyze, query, and search.   
Agentic AI is here to stay, our framework helps in shaping its future in multimedia analytics.

\hide{
\subsection{Case Study: Investigating MH17}
\review{Using only one case study to verify the framework makes its validity rather weak.
                Moreover, there is a lack of corresponding pictures for the
                description of the case, and the description is rather vague, making
                it difficult to verify the validity of the framework. More and more
                diverse validations are needed.
}
\solution{Difficult to improve when not making an actual system.}

\review{However, the current case study is rather thin; it only offers some
                suggestions for modifications from the framework perspective, without
                providing a systematic approach to using this framework. I recommend
                that the authors describe the evaluation through a more detailed case
                study.}
\solution{We should make sure that the case study focus more on what the framework brings us in terms of structuring and future improvements. The case study should start with some protocol for assessing case studies and then we should follow this for 2 or 3 case studies. Also make sure that we show explicit examples of prompt and interaction templates.}

\mf{Marcel: how about we use the rather practical idea from WEAVE, i.e. something akin to the following, but extend this with concrete references to the figures and the components and the interaction channels specifically. \newline
As an example, we now consider the investigation into the \href{https://www.government.nl/topics/mh17-incident}{flight MH17} incident, in particular the 14,579 images captured at the incident site. Many pictures show previously unseen or one-of-a-kind content for instance debris configurations and unique forensic details.
Classically, one would have used supervised classification approaches, which, however, fail to capture these categories.  Gisolf et al.~\cite{Gisolf2021mmm} proposed different strategies for interactive clustering based on deep net based visual features and a targeted clustering method. The analyst either iteratively refines by confirming subset relevance and re-adjusting the clustering threshold or querying based on image examples. With prompt-based AI alongside the ELIZA principle, the analyst has to jump between text-based and multi-media based UI components and filtering and restrictions are not necessarily in sync.
Our model suggests several ways to guide future generations of a system for such investigations, focusing on human-AI direct collaboration and UI manipulation:
It shifts the role of the user from steering all steps to giving explicit feedback on the trajectory taken.
TODO
add: Goal setting /implicit; structured result in the form of filtering; analysis, vision understanding, generation possible causes, reference to knowledge, etc (fig.2, 3)
TODO
We start with an overview of all around 10k MH17 images, alongside reports, black-box recordings, and other data sources. Looking at the flight data recording, there seems to be an abrupt loss of values visible to the analyst. The analyst queries the printout to a vision language model with the task of interpreting it. The models describes abnormal values for the last 2-3 seconds and proposes four possible main reasons (electrical outage, impact event, structural failure due to fatigue or an external explosive event). The human tells the agent to explore the structural failure hypothesis; the agent controls the image selection UI component to filter for image of bent metal using the visual analytics grammar. It clusters the images based on its understanding, and some images contain holes in the fuselage. The agent has noted these holes and annotated them, describing they are telltale signs of an explosive event and depending on the bend direction either an internal (e.g. bomb, outwards) or an external event (shrapnel, bent inwards) are possible causes. The analyst asks the agent or find similar images to this one, and in the result one is listed with much higher resolution. The human asks the AI what is visible there and where the holes come from, saying they are inward bent. the agent this could be an uncontained engine failure or a shrapnel from a SAM and it is 95\% sure this is part of the accident, i.e. neither a pre-existing fault nor impact damage, explaining why; It also queryied the web for reference images for both scenarios, displaying them to the analyst. human: show me images from the engines; engines are intact; so it is a SAM or another external explosion. Now do explosive residue testing.
Finally, incident investigation is a complex problem involving many information sources. The combination of explicit decomposition and aggregation in our model would make integral investigation of these source a future possibility, while providing a provenance and interaction trace of the analysis process. 
TODO: this needs usage of terms from our figures, the evaluation approach from Marcel, and a rewrite, this is just a very rough draft idea TODO}

As an example, we now consider the investigation into the \href{https://www.government.nl/topics/mh17-incident}{flight MH17} incident, in particular the 14,579 images captured at the incident site. Many pictures show previously unseen or one-of-a-kind content for instance debris configurations and unique forensic details.
Traditional supervised classification approaches fail to capture these categories. Gisolf et al.~\cite{Gisolf2021mmm} proposed different strategies for interactive clustering based on deep net based visual features and a targeted clustering method. The analyst either iteratively refines by confirming subset relevance and re-adjusting the clustering threshold or querying based on image examples. Evaluation with simulated users and real investigators defined the most effective strategy.

Our model suggests several ways to guide future generations of a system for such investigations. Where early deep net models were typically based on a fixed vocabulary, FMs being based on web data have a much larger vocabulary, and hence have seen far more categories already. The VA agent could therefore already assess the semantic purity of clusters and make recommendations for cluster labels. It shifts the role of the user from steering all steps to giving explicit feedback on the visualized results. By making the goal explicit ``label every image with an appropriate label'', the VA agent can use both implicit and explicit feedback to make the right recommendations at every step in the interaction sequence. In the early days, seeking feedback would use active learning to ask the user to label difficult cases. This is a notoriously hard problem as these are non-typical examples. Based on the embedded VA knowledge, the agent could now use a similar approach, but accompanied by optimal visualizations to guide the user. By making explicit why labels are suggested for a cluster, trust in the system will improve. Finally, incident investigation is a complex problem involving many information sources. The combination of explicit decomposition and aggregation in our model would make integral investigation of these source a future possibility. 

\hide{The clustering method shows an interplay between the broader (explorative) analysis tasks (clustering, re-clustering, threshold tuning) and the search and query loops (identifying smaller sets of interest from the global pool). Analysts effectively take on a human-AI teaming role: human expertise specifies items of forensic significance, while the system groups visually similar images via CNN embeddings, automatically proposing additional visually similar candidates. This reduces the amount of time spent wading through thousands of photos.
In the context of our model, this use case employs domain specific data, automatically revising thresholds for clustering granularity or dropping entire subclusters based on search and query input from the VA Agents. This iterative realignment ensures evolving investigative goals (e.g., focusing on cargo door fragments or personal items) are systematically integrated into the search hypothesis, while the observations and more detailed specifications lead to a more precise rationale.}

\hide{An obvious extension of the work would be to use relevance feedback with users indicating positive and negative image examples to learn a new classifier. Our model suggests that we should build a visual analytics agent to do this which in addition knows how to use and visualize clusters and could give recommendations. }

\subsection{\todo{Case study 2}}
}
\hide{\section{Conclusion}
We have entered a new era in which the usage of FMs and AI agents are already becoming increasingly important in any analytics solution.
In response, this paper proposes a new visual analytics model that extends several established frameworks and is explicitly designed for this new reality.

Our model addresses the emerging abilities of agentic AI (like self-reliance, inherent world knowledge, and goal strategization)--it acting more like an equal partner instead of a well-defined algorithmic process--as well as the underexplored aspect and formalization of Human-AI teaming in VA models on this more semantic- and collaborative level.
We conceptualize this through a separation in dedicated visual analytics agents that \emph{execute a strategy} through a set of prompt templates to which AI models respond with outputs and \emph{rationale} how they are obtained. In turn, the human-AI teaming component has dedicated visualizations to present the results and rationale to users, ensuring transparency and interpretability and thus building trust. The communication between humans and AI models occurs through interactive interfaces that actively let users and agents \emph{learn} how to effectively collaborate and \emph{align} on shared human values.

Naturally, our framework is \emph{one} proposal to describing an agentic VA era, which is not yet entirely filled by a single system, while also being difficult to evaluate quantitatively. We see potential for investigating, potentially domain-specific, adaptions and clarifications further.
Also, probing contracts for the proposed templates and channels, quantifying alignment dynamics, training actual reward models for VA guidance, and exploring guard-railing techniques are worthwhile endeavors.
Our work, therefore, opens the discussion on promising research directions for the wider scientific community, thereby providing a foundation for future research practices in human-AI collaboration.}

\clearpage
\bibliographystyle{ACM-Reference-Format}
\bibliography{template}
\end{document}

%% file: packages.tex
\usepackage[inkscapelatex=false]{svg}
\usepackage{tikz}
\usetikzlibrary{calc,patterns,patterns.meta}
\usepackage{enumitem}
\usepackage{changepage}
\usepackage[most]{tcolorbox}
\usepackage{tabu}                      
\usepackage{booktabs}                  
\usepackage{tabularx}                  
\usepackage{array}
\usepackage{mwe}                       

\usepackage{mathptmx}                  
\usepackage{microtype}

%% file: commands.tex
\newcommand{\coloredtextsymbol}[2][]{%
  \tikz[baseline=(n.base)]%
    \node[
      inner sep=2pt,
      outer sep=0pt,
      rounded corners=2pt,
      draw=none,
      fill=black!20,
      font=\sffamily\footnotesize,
      #1
    ] (n) {#2};%
}

\colorlet{ColorSymbolDesignBG}{red!15}
\definecolor{ColorSymbolDesignText}{HTML}{501818}

\newcommand{\symbolDesignNo}[1]{\texorpdfstring{\coloredtextsymbol[fill=ColorSymbolDesignBG, text=ColorSymbolDesignText]{#1}}{}}

\newif\ifshowcomments
\showcommentstrue

\newcommand{\review}[1] {
\ifshowcomments
    \textcolor{red}{\\ \noindent REVIEWER: #1 \\}
\else
    {}
\fi}

\newcommand{\mf}[1]{
\ifshowcomments
    \textcolor{blue}{\\ MF: #1 \\}
\else
    {}
\fi}

\newcommand{\todo}[1]{
\ifshowcomments
    \textcolor{red}{TODO #1}
\else
    {}
\fi}

\newcommand{\solution}[1]{
\ifshowcomments
    \textcolor{blue}{\noindent SOLUTION #1 \\}
\else
    {}
\fi}

\newcommand{\parheader}[1]{\textbf{#1} --- }

\newcommand{\hide}[1]{{}}

\newcommand{\conceptdefinition}[2]
{
\begin{adjustwidth}{2mm}{0mm}
    \textbf{#1:} #2
\end{adjustwidth}
}

\newcommand{\Dconsist}{\symbolDesignNo{D1}}
\newcommand{\Dfunction}{\symbolDesignNo{D2}}
\newcommand{\Dlimit}{\symbolDesignNo{D3}}
\newcommand{\Dshift}{\symbolDesignNo{D6}}
\newcommand{\Dinteract}{\symbolDesignNo{D5}}
\newcommand{\Dvisualize}{\symbolDesignNo{D4}}

\definecolor{ColorAI}{RGB}{212, 235, 193}
\definecolor{ColorAIAgent}{RGB}{81, 119, 52}
\definecolor{ColorHumanAITeaming}{RGB}{244, 228, 216}
\definecolor{ColorVAAgent}{RGB}{155, 39, 31}
\definecolor{ColorHumanUnderstanding}{RGB}{248, 244, 216}
\definecolor{ColorExpert}{RGB}{175, 152, 53}
\definecolor{ColorChannel}{RGB}{192, 191, 191}
\definecolor{ColorAction}{RGB}{206, 119, 66}

\newtcbox{\bgcolorbox}[1][yellow]{on line,
  colback=#1,  
  boxrule=0pt,
  arc=2pt,
  left=-1pt,right=-1pt,top=-2pt,bottom=-2pt}

\newtcbox{\bgcolorboxstriped}[1][yellow]{on line,
  colback=#1,   
  boxrule=0pt,
  arc=2pt,
  left=-1pt, right=-1pt, top=-2pt, bottom=-2pt,
  enhanced,
  overlay={%
    \fill[
    pattern={Lines[angle=45,distance=12pt, line width=6pt]},
    pattern color=ColorAction!15,
    ]
      (interior.south west) rectangle (interior.north east);
  }
}

\graphicspath{{figs/}{figures/}{pictures/}{images/}{./}} 

\newcolumntype{L}{>{\raggedright\arraybackslash}X}

%% file: template.bib
@string{TVCG="IEEE TVCG"}

@string{MMM="MMM"}

@string{CGA="IEEE CG\&A"}

@string{ICML="ICML"}

@string{CGF="Computer Graphics Forum"}

@string{CGF="CGF"}

@string{PAMI="IEEE PAMI"}

@string{ACMMM="ACM Multimedia"}

@ARTICLE{Maciej_graph_of_thought2024,
  author={Besta, Maciej and Memedi, Florim and Zhang, Zhenyu and Gerstenberger, Robert and Piao, Guangyuan and Blach, Nils and others},
  journal=PAMI, 
  title={Demystifying Chains, Trees, and Graphs of Thoughts}, 
  year={2025},
  volume={47},
  number={12},
  pages={10967-10989},
  doi={10.1109/TPAMI.2025.3598182}}

@article{chinchor2010multimedia,
  title={Multimedia analysis+ visual analytics= multimedia analytics},
  author={Chinchor, Nancy A and Thomas, James J and Wong, Pak Chung and Christel, Michael G and Ribarsky, William},
  journal=CGA,
  volume={30},
  number={5},
  pages={52--60},
  year={2010},
  doi = {10.1109/MCG.2010.92},
  publisher={IEEE}
}

@misc{AndrienkoATWL2025,
    title = {ATWL: A Formal Language for Representing, Comparing, and Reusing Visual Analytics Workflows},
    author = {Natalia Andrienko and Gennady Andrienko and Jürgen Bernard and Michael Sedlmair},
    eprint = {2605.25489},
    archivePrefix = {arXiv},
    url={https://arxiv.org/abs/2605.25489}, 
    year = 2026
}

@article{Huang2025hallucination,
   title={A Survey on Hallucination in Large Language Models: Principles, Taxonomy, Challenges, and Open Questions},
   volume={43},
   ISSN={1558-2868},
   url={http://dx.doi.org/10.1145/3703155},
   DOI={10.1145/3703155},
   number={2},
   journal={ACM Trans. on Inform. Systems},
   publisher={ACM},
   author={Huang, Lei and Yu, Weijiang and Ma, Weitao and Zhong, Weihong and Feng, Zhangyin and Wang, Haotian and Chen, Qianglong and Peng, Weihua and Feng, Xiaocheng and Qin, Bing and Liu, Ting},
   year={2025},
   pages={1–55} }

@misc{chen2025interchat,
      title={InterChat: Enhancing Generative Visual Analytics using Multimodal Interactions}, 
      author={Juntong Chen and Jiang Wu and Jiajing Guo and Vikram Mohanty and Xueming Li and Jorge Piazentin Ono and Wenbin He and Liu Ren and Dongyu Liu},
      year={2025},
      eprint={2503.04110},
      archivePrefix={arXiv},
      primaryClass={cs.HC},
      url={https://arxiv.org/abs/2503.04110}, 
      doi = {10.48550/arXiv.2503.04110}
}

@ARTICLE{Li2023Attention,
author={Li, Yiran and Wang, Junpeng and Dai, Xin and Wang, Liang and Yeh, Chin-Chia Michael and Zheng, Yan and Zhang, Wei and Ma, Kwan-Liu},
journal=TVCG,
title={{ How Does Attention Work in Vision Transformers? A Visual Analytics Attempt }},
year={2023},
volume={29},
number={06},
ISSN={1941-0506},
pages={2888-2900},
doi={10.1109/TVCG.2023.3261935},
url = {https://doi.ieeecomputersociety.org/10.1109/TVCG.2023.3261935},
publisher={IEEE Computer Society}
}

@misc{karpas2022mrklsystemsmodularneurosymbolic,
      title={{MRKL} Systems: A modular, neuro-symbolic architecture that combines large language models, external knowledge sources and discrete reasoning}, 
      author={Ehud Karpas and others},
      year={2022},
      eprint={2205.00445},
      archivePrefix={arXiv},
      primaryClass={cs.CL},
      url={https://arxiv.org/abs/2205.00445}, 
      doi = {10.48550/arXiv.2205.00445}
}

@inproceedings{Guo2024MultiAgentSurvey890,
  title     = {Large Language Model Based Multi-agents: A Survey of Progress and Challenges},
  author    = {Guo, Taicheng and Chen, Xiuying and Wang, Yaqi and Chang, Ruidi and Pei, Shichao and Chawla, Nitesh V. and Wiest, Olaf and Zhang, Xiangliang},
  booktitle = {IJCAI},
  year      = {2024},
  doi       = {10.24963/ijcai.2024/890},
  url       = {https://doi.org/10.24963/ijcai.2024/890},
}

@article{Xi2025LLMagentsurvey,
    author = {Xi, Z. and Chen, W. and Guo, X. and others},
    title = {The rise and potential of large language model based agents: a survey},
    journal = {Science China Information Sciences},
    volume = {68},
    issue = {2},
    doi = {10.1007/s11432-024-4222-0},
    year = {2025}
}

@INPROCEEDINGS {Lin2024evaluatingGenAI,
author = { Lin, Tica and Pfister, Hanspeter and Wang, Jui-Hsien },
booktitle = { 2024 IEEE 17th Pacific Visualization Conference (PacificVis) },
title = {{ GenLens: A Systematic Evaluation of Visual {GenAI} Model Outputs }},
year = {2024},
volume = {},
ISSN = {},
pages = {313-318},
doi = {10.1109/PacificVis60374.2024.00044},
url = {https://doi.ieeecomputersociety.org/10.1109/PacificVis60374.2024.00044},
publisher = {IEEE Computer Society},
month =apr}

@misc{schulhoff2025promptreportsystematicsurvey,
      title={The Prompt Report: A Systematic Survey of Prompt Engineering Techniques}, 
      author={Sander Schulhoff and others},
      year={2025},
      eprint={2406.06608},
      archivePrefix={arXiv},
      primaryClass={cs.CL},
      url={https://arxiv.org/abs/2406.06608},
      doi = {10.48550/arXiv.2406.06608}
}

@article{weizenbaum1966eliza,
  title={{ELIZA}—a computer program for the study of natural language communication between man and machine},
  author={Weizenbaum, Joseph},
  journal={Communications of the ACM},
  volume={9},
  number={1},
  pages={36--45},
  year={1966},
  doi = {10.1145/365153.365168},
  publisher={ACM}
}

@misc{ji2024aialignmentcomprehensivesurvey,
      title={{AI} Alignment: A Comprehensive Survey}, 
      author={Jiaming Ji and Tianyi Qiu and Boyuan Chen and Borong Zhang and others},
      year={2024},
      eprint={2310.19852},
      archivePrefix={arXiv},
      primaryClass={cs.AI},
      url={https://arxiv.org/abs/2310.19852}, 
      doi = {10.48550/arXiv.2310.19852}
}

@article{north2006toward,
  title={Toward measuring visualization insight},
  author={North, Chris},
  journal=CGA,
  volume={26},
  number={3},
  pages={6--9},
  year={2006},
  publisher={IEEE}
}

@book{keim2008visual,
  title={Visual analytics: Definition, process, and challenges},
  author={Keim, Daniel and Andrienko, Gennady and Fekete, Jean-Daniel and G{\"o}rg, Carsten and Kohlhammer, J{\"o}rn and Melan{\c{c}}on, Guy},
  year={2008},
  publisher={Springer}
}

@article{mosqueira2023human,
  title={Human-in-the-loop machine learning: a state of the art},
  author={Mosqueira-Rey, Eduardo and Hern{\'a}ndez-Pereira, Elena and Alonso-R{\'\i}os, David and Bobes-Bascar{\'a}n, Jos{\'e} and Fern{\'a}ndez-Leal, {\'A}ngel},
  journal={A. I. Review},
  volume={56},
  number={4},
  pages={3005--3054},
  year={2023},
  publisher={Springer}
}

@inproceedings{sacha2016human,
  title={Human-centered machine learning through interactive visualization},
  author={Sacha, Dominik and Sedlmair, Michael and Zhang, Leishi and Lee, J and Weiskopf, Daniel and North, Stephen and Keim, Daniel},
  booktitle={24th ESANN},
  year={2016}
}

@article{retzlaff2024human,
  title={Human-in-the-loop reinforcement learning: A survey and position on requirements, challenges, and opportunities},
  author={Retzlaff, Carl and others},
  journal={Journal of Artificial Intelligence Research},
  volume={79},
  pages={359--415},
  year={2024}
}

@article{jiang2019recent,
  title={Recent research advances on interactive machine learning},
  author={Jiang, Liu and Liu, Shixia and Chen, Changjian},
  journal={Journal of Visualization},
  volume={22},
  pages={401--417},
  year={2019},
  publisher={Springer}
}

@article{vaswani2017attention,
  title={Attention is all you need},
  author={Vaswani, A},
  journal={NeurIPS},
  year={2017}
}

@inproceedings{zahalka2014towards,
  title={Towards interactive, intelligent, and integrated multimedia analytics},
  author={Zah{\'a}lka, Jan and Worring, Marcel},
  booktitle={IEEE VAST},
  year={2014},
  doi={10.1109/VAST.2014.7042476}
}

@ARTICLE{worring2016pivot,
  author={Worring, Marcel and Koelma, Dennis and Zahálka, Jan},
  journal={IEEE Transactions on Multimedia}, 
  title={Multimedia Pivot Tables for Multimedia Analytics on Image Collections}, 
  year={2016},
  volume={18},
  number={11},
  pages={2217-2227},
  doi={10.1109/TMM.2016.2614380}}

@ARTICLE{sacha2014knowlegde,
  author={Sacha, Dominik and Stoffel, Andreas and Stoffel, Florian and Kwon, Bum Chul and Ellis, Geoffrey and Keim, Daniel A.},
  journal=TVCG, 
  series = {TVCG},
  title={Knowledge Generation Model for Visual Analytics}, 
  year={2014},
  volume={20},
  number={12},
  pages={1604-1613},
  doi={10.1109/TVCG.2014.2346481}}

@article{perez2022typology,
  title={A typology of guidance tasks in mixed-initiative visual analytics environments},
  author={P{\'e}rez-Messina, Ignacio and Ceneda, Davide and El-Assady, Mennatallah and Miksch, Silvia and Sperrle, Fabian},
  journal=CGF,
  volume={41},
  number={3},
  pages={465--476},
  year={2022},
  organization={Wiley Online Library},
  doi = {10.1111/cgf.14555}
}

@article{Sperrle.CoAdaptiveGuidance.2021,
author = {Sperrle, F. and Schäfer, H. and Keim, D. and El-Assady, M.},
title = {Learning Contextualized User Preferences for Co-Adaptive Guidance in Mixed-Initiative Topic Model Refinement},
journal = {CGF},
volume = {40},
number = {3},
pages = {215-226},
doi = {https://doi.org/10.1111/cgf.14301},
year = {2021}
}

@inproceedings{MorrisAGI2024,
author = {Morris, Meredith and others},
title = {Position: levels of AGI for operationalizing progress on the path to AGI},
year = {2024},
booktitle = {ICML}
}

@inproceedings{Wang2020,
author = {Wang, Dakuo and Churchill, Elizabeth and Maes, Pattie and Fan, Xiangmin and Shneiderman, Ben and Shi, Yuanchun and Wang, Qianying},
title = {From Human-Human Collaboration to Human-{AI} Collaboration: Designing {AI} Systems That Can Work Together with People},
year = {2020},
doi = {10.1145/3334480.3381069},
booktitle = {CHI EA '20},
pages = {1–6},
}

@misc{metz2025mappingspacehumanfeedback,
      title={Mapping out the Space of Human Feedback for Reinforcement Learning: A Conceptual Framework}, 
      author={Yannick Metz and David Lindner and Raphaël Baur and Mennatallah El-Assady},
      year={2025},
      eprint={2411.11761},
      archivePrefix={arXiv},
      primaryClass={cs.LG},
      doi={10.48550/arXiv.2411.11761}, 
}

@ARTICLE{Feng2024promptmagician,
author={Feng, Yingchaojie and Wang, Xingbo and Wong, Kam Kwai and Wang, Sijia and Lu, Yuhong and Zhu, Minfeng and Wang, Baicheng and Chen, Wei},
journal=TVCG,
title={{ PromptMagician: Interactive Prompt Engineering for Text-to-Image Creation }},
year={2024},
volume={30},
number={01},
pages={295-305},
doi={10.1109/TVCG.2023.3327168}
}

@misc{casper2025aiagentindex,
      title={The {AI} Agent Index}, 
      author={Stephen Casper and others},
      year={2025},
      eprint={2502.01635},
      archivePrefix={arXiv},
      primaryClass={cs.SE},
      doi={10.48550/arXiv.2502.01635}, 
}

@inproceedings{gia2025vbs,
author = {Gia, Bao Tran and Khanh, Tuong Bui Cong and Thanh, Tam Le Thi and Doan, Thuyen Tran and Le, Khiem and Do, Tien and Mai, Tien-Dung and Ngo, Thanh Duc and Le, Duy-Dinh and Satoh, Shin’ichi},
title = {{NII-UIT} at {VBS2025}: Multimodal Video Retrieval with LLM Integration and Dynamic Temporal Search},
year = {2025},
doi = {10.1007/978-981-96-2074-6_38},
booktitle = MMM

}

@inproceedings{zahalka2015analytic,
author = {Zah\'{a}lka, Jan and Rudinac, Stevan and Worring, Marcel},
title = {Analytic Quality: Evaluation of Performance and Insight in Multimedia Collection Analysis},
booktitle = {ACM Multimedia},
year = {2015},
url = {https://doi.org/10.1145/2733373.2806279},
doi = {10.1145/2733373.2806279},
location = {Brisbane, Australia}
}

@inproceedings{Gisolf2021mmm,
author = {Gisolf, Floris and Geradts, Zeno and Worring, Marcel},
title = {Search and Explore Strategies for Interactive Analysis of Real-Life Image Collections with Unknown and Unique Categories},
booktitle = MMM,
year = {2021},
doi = {10.1007/978-3-030-67835-7_21}
}

@inproceedings{Gou2024CRITIC,
    author = {Zhibin Gou and Zhihong Shao and Yeyun Gong and Yelong Shen
and Yujiu Yang and Nan Duan and Weizhu Chen}, 
    title = {{CRITIC}, Large language models can self-correct
with tool-interactive critiquing},
    booktitle = {ICLR},
    year = {2024}
}

@ARTICLE{Tang2022VideoModerator,
  author={Tang, Tan and Wu, Yanhong and Wu, Yingcai and Yu, Lingyun and Li, Yuhong},
  journal=TVCG, 
  title={VideoModerator: A Risk-aware Framework for Multimodal Video Moderation in E-Commerce}, 
  year={2022},
  volume={28},
  number={1},
  pages={846-856},
  doi={10.1109/TVCG.2021.3114781}
}

@article{fabbrizzi2022bias,
  title={A survey on bias in visual datasets},
  author={Fabbrizzi, Simone and Papadopoulos, Symeon and Ntoutsi, Eirini and Kompatsiaris, Ioannis},
  journal={Computer Vision and Image Understanding},
  volume={223},
  pages={103552},
  year={2022},
  doi = {10.1016/j.cviu.2022.103552},
  publisher={Elsevier}
}

@INPROCEEDINGS{Kokoshka2014climate,
  author={Kokoschka, Vanessa and Secco, Cristian A. and Nazemi, Kawa},
  booktitle={28th Int. Conf. Information Visualisation (IV)}, 
  title={Visual Analytics - Climate Change in Social Media}, 
  year={2024},
  volume={},
  number={},
  pages={167-173},
  doi={10.1109/IV64223.2024.00037}
}

@INPROCEEDINGS{Bannach2017radiomics,
  author={Bannach, Andreas and Bernard, Jürgen and Jung, Florian and Kohlhammer, Jörn and May, Thorsten and Scheckenbach, Kathrin and Wesarg, Stefan},
  booktitle={2017 IEEE Workshop on VAHC}, 
  title={Visual analytics for radiomics: Combining medical imaging with patient data for clinical research}, 
  year={2017},
  volume={},
  number={},
  pages={84-91},
  doi={10.1109/VAHC.2017.8387545}
}

@article{Lokuc2023VBS,
    author = {Loko\v{c}, J. and Andreadis, S. and Bailer, W. and others},
    title = {Interactive video retrieval in the age of effective joint embedding deep models: lessons from the 11th {VBS}},
    journal = {Multimedia Systems},
    volume = {29},
    pages = {3481–3504},
    year = {2023},
    doi = {10.1007/s00530-023-01143-5}
}

@article{Rietveld2020Instagram,
author = {Robert Rietveld and Willemijn Van Dolen and Masoud Mazloom and Marcel Worring},
title ={What you Feel, Is what you like Influence of Message Appeals on Customer Engagement on Instagram},
journal = {Journal of Interactive Marketing},
volume = {49},
number = {1},
pages = {20-53},
year = {2020},
doi = {10.1016/j.intmar.2019.06.003}
}

@inproceedings{Fa2024RAGmeetingLLM,
author = {Fan, Wenqi and Ding, Yujuan and Ning, Liangbo and Wang, Shijie and Li, Hengyun and Yin, Dawei and Chua, Tat-Seng and Li, Qing},
title = {A Survey on RAG Meeting LLMs: Towards Retrieval-Augmented Large Language Models},
year = {2024},
url = {https://doi.org/10.1145/3637528.3671470},
doi = {10.1145/3637528.3671470},
booktitle = {ACM SIGKDD}
}

@inproceedings{Patrick2020RAGforKnowledgeIntensive,
author = {Lewis, Patrick and others},
title = {Retrieval-augmented generation for knowledge-intensive {NLP} tasks},
year = {2020},
booktitle = {NIPS},
}

@inproceedings{Yasunga2023RAGMultimodal,
author = {Yasunaga, Michihiro and Aghajanyan, Armen and Shi, Weijia and James, Rich and Leskovec, Jure and Liang, Percy and Lewis, Mike and Zettlemoyer, Luke and Yih, Wen-tau},
title = {Retrieval-augmented multimodal language modeling},
year = {2023},
booktitle = ICML
}

@inproceedings{zhao2023-retrieving,
    title = "Retrieving Multimodal Information for Augmented Generation: A Survey",
    author = "Zhao, Ruochen  and
      others",
    booktitle = "EMNLP",
    year = "2023",
    doi = "10.18653/v1/2023.findings-emnlp.314"
}

@article{fu2023more,
  title={More than data stories: Broadening the role of visualization in contemporary journalism},
  author={Fu, Yu and Stasko, John},
  journal=TVCG,
  year={2023},
  publisher={IEEE}
}

@book{keim2010mastering,
  title={Mastering the information age solving problems with visual analytics},
  author={Keim, Daniel and Kohlhammer, J{\"o}rn and Ellis, Geoffrey and Mansmann, Florian},
  year={2010},
  publisher={Eurographics Association}
}

@article{Andrienko.ModelBuilding.2018,
author = {Andrienko, N. and Lammarsch, T. and Andrienko, G. and Fuchs, G. and Keim, D. and Miksch, S. and Rind, A.},
title = {Viewing Visual Analytics as Model Building},
journal = CGF,
volume = {37},
number = {6},
pages = {275-299},
doi = {https://doi.org/10.1111/cgf.13324},
url = {https://onlinelibrary.wiley.com/doi/abs/10.1111/cgf.13324},
eprint = {https://onlinelibrary.wiley.com/doi/pdf/10.1111/cgf.13324},
year = {2018}
}

@INPROCEEDINGS{Green.VAHumanCognition.2008,
  author={Green, Tera Marie and Ribarsky, William and Fisher, Brian},
  booktitle={VAST}, 
  title={Visual analytics for complex concepts using a human cognition model}, 
  year={2008},
  doi={10.1109/VAST.2008.4677361}}

@article{Green.HumanCognitionModelVA.2009,
  title = {Building and Applying a Human Cognition Model for Visual Analytics},
  volume = {8},
  ISSN = {1473-8724},
  url = {http://dx.doi.org/10.1057/ivs.2008.28},
  DOI = {10.1057/ivs.2008.28},
  number = {1},
  journal = {Information Visualization},
  publisher = {SAGE Publications},
  author = {Green,  Tera Marie and Ribarsky,  William and Fisher,  Brian},
  year = {2009},
  pages = {1–13}
}

@inproceedings{FiHiJe2022VAIntelligence,
 author = {Fischer, Maximilian T. and Hirsbrunner, Simon D. and Jentner, Wolfgang and Miller, Matthias and Keim, Daniel A. and Helm, Paula},
 booktitle = {FAccT},
 doi = {10.1145/3531146.3533151},
 isbn = {978-1-4503-9352-2},
 title = {Promoting Ethical Awareness in Communication Analysis: Investigating Potentials and Limits of Visual Analytics for Intelligence Applications},
 year = {2022}
}

@book{munzner2015visualization,
  author = {Munzner, T.},
  isbn = {9781498759717},
  publisher = {CRC Press},
  title = {Visualization Analysis and Design},
  year = 2015
}

@article{Ulmer2024Progressive,
  author={Ulmer, Alex and Angelini, Marco and Fekete, Jean-Daniel and Kohlhammer, Jorn and May, Thorsten},
  journal=TVCG,
  title={{ A Survey on Progressive Visualization }},
  year={2024},
  volume={30},
  number={09},
  ISSN={1941-0506},
  pages={6447-6467},
  doi={10.1109/TVCG.2023.3346641},
  publisher={IEEE Computer Society}
}

@misc{Fischer.MultiCase.2024,
      title={MULTI-CASE: A Transformer-based Ethics-aware Multimodal Investigative Intelligence Framework}, 
      author={Maximilian T. Fischer and Yannick Metz and Lucas Joos and Matthias Miller and Daniel A. Keim},
      year={2024},
      archivePrefix={arXiv},
      doi={10.48550/arXiv.2401.01955}, 
}

@article{zahalka2017blackthorn,
  title={Blackthorn: large-scale interactive multimodal learning},
  author={Zah{\'a}lka, Jan and Rudinac, Stevan and J{\'o}nsson, Bj{\"o}rn {\TH}{\'o}r and Koelma, Dennis C and Worring, Marcel},
  journal={IEEE Transactions on Multimedia},
  volume={20},
  number={3},
  pages={687--698},
  year={2017}
}

@inproceedings{khan2020interactive,
  title={Interactive learning for multimedia at large},
  author={Khan, Omar Shahbaz and J{\'o}nsson, Bj{\"o}rn {\TH}{\'o}r and Rudinac, Stevan and Zah{\'a}lka, Jan and Ragnarsd{\'o}ttir, Hanna and {\TH}orleiksd{\'o}ttir, {\TH}{\'o}rhildur and Gu{\dh}mundsson, Gylfi {\TH}{\'o}r and Amsaleg, Laurent and Worring, Marcel},
  booktitle= {ECIR}, 
  year={2020}
}

@inproceedings{khan2024exquisitor,
  title={Exquisitor: Studying the Interplay Between Conversational Search and Relevance Feedback},
  author={Khan, Omar Shahbaz and Sharma, Ujjwal and Rudinac, Stevan and J{\'o}nsson, Bj{\"o}rn {\TH}{\'o}r},
  booktitle={CBMI},
  year={2024}
}

@article{zahalka2015interactive,
  title={Interactive multimodal learning for venue recommendation},
  author={Zah{\'a}lka, Jan and Rudinac, Stevan and Worring, Marcel},
  journal={IEEE Transactions on Multimedia},
  volume={17},
  number={12},
  pages={2235--2244},
  year={2015},
  publisher={IEEE}
}

@article{he2023videopro,
  title={VideoPro: A visual analytics approach for interactive video programming},
  author={He, Jianben and Wang, Xingbo and Wong, Kam Kwai and Huang, Xijie and Chen, Changjian and Chen, Zixin and Wang, Fengjie and Zhu, Min and Qu, Huamin},
  journal=TVCG,
  volume={30},
  number={1},
  pages={87--97},
  year={2023},
  publisher={IEEE}
}

@article{batch2023uxsense,
  title={{uxSense}: Supporting user experience analysis with visualization and computer vision},
  author={Batch, Andrea and Ji, Yipeng and Fan, Mingming and Zhao, Jian and Elmqvist, Niklas},
  journal=TVCG,
  year={2023},
  publisher={IEEE}
}

@article{yuan2023rl,
  title={Rl-vigen: A reinforcement learning benchmark for visual generalization},
  author={Yuan, Zhecheng and Yang, Sizhe and Hua, Pu and Chang, Can and Hu, Kaizhe and Xu, Huazhe},
  journal={NeurIPS},
  volume={36},
  pages={6720--6747},
  year={2023}
}

@article{ji2023safety,
  title={Safety gymnasium: A unified safe reinforcement learning benchmark},
  author={Ji, Jiaming and Zhang, Borong and Zhou, Jiayi and Pan, Xuehai and Huang, Weidong and Sun, Ruiyang and Geng, Yiran and Zhong, Yifan and Dai, Josef and Yang, Yaodong},
  journal={NeurIPS},
  volume={36},
  pages={18964--18993},
  year={2023}
}

@misc{lin2023mcu,
  title={Mcu: A task-centric framework for open-ended agent evaluation in minecraft},
  author={Lin, Haowei and Wang, Zihao and Ma, Jianzhu and Liang, Yitao},
  doi={10.48550/arXiv.2310.08367},
  year={2023}
}

@incollection{Hart1988NasaTLX,
  title = {Development of NASA-TLX (Task Load Index): Results of Empirical and Theoretical Research},
  series = {Advances in Psychology},
  volume = {52},
  pages = {139--183},
  year = {1988},
doi = {10.1016/S0166-4115(08)62386-9},
  booktitle = {Human Mental Workload},
  author = {Sandra G. Hart and Lowell E. Staveland},
  publisher = {North-Holland},
}

@inproceedings{zhao2021calibrate,
  title={Calibrate before use: Improving few-shot performance of language models},
  author={Zhao, Zihao and Wallace, Eric and Feng, Shi and Klein, Dan and Singh, Sameer},
  booktitle={International conference on machine learning},
  pages={12697--12706},
  year={2021},
  organization={PMLR}
}

@article{liu2024lost,
  title={Lost in the middle: How language models use long contexts},
  author={Liu, Nelson F and Lin, Kevin and Hewitt, John and Paranjape, Ashwin and Bevilacqua, Michele and Petroni, Fabio and Liang, Percy},
  journal={Trans. Assoc. for Computational Linguistics},
  volume={12},
  pages={157--173},
  year={2024},
  publisher={MIT Press}
}

@article{Hutchinson.FoundationalVAOpportunities.2025,
title = {Foundation model assisted visual analytics: Opportunities and Challenges},
journal = {Computers \& Graphics},
volume = {130},
pages = {104246},
year = {2025},
issn = {0097-8493},
doi = {https://doi.org/10.1016/j.cag.2025.104246},
author = {Maeve Hutchinson and Radu Jianu and Aidan Slingsby and Pranava Madhyastha}
}

@inproceedings{Bernard.HumanDataModelVA.2025,
booktitle = {EuroVis Workshop on Visual Analytics (EuroVA)},
editor = {Schulz, Hans-Jörg and Villanova, Anna},
title = {{The Human-Data-Model Interaction Canvas for Visual Analytics}},
author = {Bernard, Jürgen},
year = {2025},
publisher = {EG},
ISSN = {2664-4487},
ISBN = {978-3-03868-283-7},
DOI = {10.2312/eurova.20251096}
}

@misc{mcp,
  author       = {Anthropic},
  title        = {Model Context Protocol ({MCP})},
  year         = {2025},
  url          = {https://modelcontextprotocol.io/docs/getting-started/intro}
}

@inproceedings{Sheng2026MultiAgent,
    author = {R. Sheng and Y. Yang and  C. Shi and Y. Lin and Z. Chen and H Qu and F. Cheng},
    title = {DiLLS: Interactive Diagnosis of LLM-based Multi-agent Systems via Layered Summary of Agent Behaviors},
    booktitle = {Proceedings of SIGCHI},
    year = 2026
}

@article{cohen2024contextcite,
  title={{ContextCite}: Attributing model generation to context},
  author={Cohen-Wang, Benjamin and Shah, Harshay and Georgiev, Kristian and Madry, Aleksander},
  journal={NeurIPS},
  volume={37},
  pages={95764--95807},
  year={2024}
}

@article{wei2022chain,
  title={Chain-of-thought prompting elicits reasoning in large language models},
  author={Wei, Jason and others},
  IGNORAUTHORS={Wei, Jason and Wang, Xuezhi and Schuurmans, Dale and Bosma, Maarten and Xia, Fei and Chi, Ed and Le, Quoc V and Zhou, Denny and others},
  journal={NeurIPS},
  volume={35},
  pages={24824--24837},
  year={2022}
}

@article{kojima2022large,
  title={Large language models are zero-shot reasoners},
  author={Kojima, Takeshi and Gu, Shixiang Shane and Reid, Machel and Matsuo, Yutaka and Iwasawa, Yusuke},
  journal={NeurIPS},
  volume={35},
  pages={22199--22213},
  year={2022}
}

@misc{vats2024surveyhumanaiteaminglarge,
  title={A Survey on Human-{AI} Teaming with Large Pre-Trained Models}, 
  author={Vanshika Vats and Marzia Binta Nizam and Minghao Liu and Ziyuan Wang and Richard Ho and others},
 IGNOREfullauthors={Vanshika Vats and Marzia Binta Nizam and Minghao Liu and Ziyuan Wang and Richard Ho and Mohnish Sai Prasad and Vincent Titterton and Sai Venkat Malreddy and Riya Aggarwal and Yanwen Xu and Lei Ding and Jay Mehta and Nathan Grinnell and Li Liu and Sijia Zhong and Devanathan Nallur Gandamani and Xinyi Tang and Rohan Ghosalkar and Celeste Shen and Rachel Shen and Nafisa Hussain and Kesav Ravichandran and James Davis},
  year={2024},
  eprint={2403.04931},
  archivePrefix={arXiv},
  primaryClass={cs.AI},
  url={https://arxiv.org/abs/2403.04931}, 
}

@misc{gisolf2025hypergraphs,
    title =  {Interactive Hypergraph Visual Analytics for Exploring Large and Complex Image Collections},
    author = {F. Gisolf and Z.J.M.H. Geradts and M. Worring},
    year = {2025}, 
    doi = {10.48550/2510.20050},
    eprint = {2510.20050},
    archivePrefix = {arXiv},
    url = {https://arxiv.org/abs/2510.20050}
}

@ARTICLE{Satyanarayan2017vega,
  author={Satyanarayan, Arvind and Moritz, Dominik and Wongsuphasawat, Kanit and Heer, Jeffrey},
  journal={IEEE TVCG}, 
  title={Vega-Lite: A Grammar of Interactive Graphics}, 
  year = {2017},
  doi={10.1109/TVCG.2016.2599030}}

@inproceedings{Palani2026Lexara,
    author = {S. Palani and V. Setlur},
    title = {Lexara: A User-Centered Toolkit for Evaluating Large Language Models for Conversational Visual Analytics},
    booktitle = {CHI},
    year = {2026}
}

@misc{fragiadakis2025evaluating,
  title={Evaluating Human-AI Collaboration: A Review and Methodological Framework}, 
  author={George Fragiadakis and Christos Diou and George Kousiouris and Mara Nikolaidou},
  year={2025},
  eprint={2407.19098},
  archivePrefix={arXiv},
  primaryClass={cs.HC},
  url={https://arxiv.org/abs/2407.19098}, 
}

@article{monadjemi2023agentbased,
  author = {Monadjemi, Shayan and Guo, Mengtian and Gotz, David and Garnett, Roman and Ottley, Alvitta},
  title = {Human–Computer Collaboration for Visual Analytics: an Agent-based Framework},
  journal = {CGF},
  volume = {42},
  number = {3},
  pages = {199--210},
  doi = {https://doi.org/10.1111/cgf.14823},
  year = {2023}
}

@article{zhao2025lightweightVA,
  author={Zhao, Yuheng and others},
IGNOREFULLAUTHORS={Zhao, Yuheng and Wang, Junjie and Xiang, Linbing and Zhang, Xiaowen and Guo, Zifei and Turkay, Cagatay and Zhang, Yu and Chen, Siming},
  journal={IEEE TVCG},
  title={{LightVA: Lightweight Visual Analytics With LLM Agent-Based Task Planning and Execution }},
  year={2025},
  volume={31},
  number={09},
  pages={6162--6177},
  doi={10.1109/TVCG.2024.3496112},
  publisher={IEEE Computer Society},
  address={Los Alamitos, CA, USA},
}

@article{schick2021self,
  title={Self-diagnosis and self-debiasing: A proposal for reducing corpus-based bias in nlp},
  author={Schick, Timo and Udupa, Sahana and Sch{\"u}tze, Hinrich},
  journal={Transactions of the Association for Computational Linguistics},
  volume={9},
  pages={1408--1424},
  year={2021},
  publisher={MIT Press One Rogers Street, Cambridge, MA 02142-1209, USA journals-info~…}
}

@article{inan2023llama,
  title={{L}lama guard: {LLM}-based input-output safeguard for human-{AI} conversations},
  author={Inan, Hakan and Upasani, Kartikeya and Chi, Jianfeng and Rungta, Rashi and Iyer, Krithika and Mao, Yuning and Tontchev, Michael and Hu, Qing and Fuller, Brian and others},
  journal={arXiv:2312.06674},
  year={2023}
}

@article{ouyang2022training,
  title={Training language models to follow instructions with human feedback},
  author={Ouyang, Long and others},
IGNOREFULLAUTHORS={Ouyang, Long and Wu, Jeffrey and Jiang, Xu and Almeida, Diogo and Wainwright, Carroll and Mishkin, Pamela and Zhang, Chong and Agarwal, Sandhini and Slama, Katarina and others},
  journal={NeurIPS},
  volume={35},
  pages={27730--27744},
  year={2022}
}

@article{stiennon2020learning,
  title={Learning to summarize with human feedback},
  author={Stiennon, Nisan and Ouyang, Long and Wu, Jeffrey and Ziegler, Daniel and Lowe, Ryan and Voss, Chelsea and Radford, Alec and Amodei, Dario and Christiano, Paul F},
  journal={NeurIPS},
  volume={33}, 
  year = {2020},
  pages={3008--3021},
}

@article{wang2022exploring,
  title={Exploring the limits of domain-adaptive training for detoxifying large-scale language models},
  author={Wang, Boxin and others},
  IGNOREFULLAUTHORS={Wang, Boxin and Ping, Wei and Xiao, Chaowei and Xu, Peng and Patwary, Mostofa and Shoeybi, Mohammad and Li, Bo and Anandkumar, Anima and Catanzaro, Bryan},
  journal={NeurIPS},
  volume={35},
  pages={35811--35824},
  year={2022}
}

@article{Yi2007interaction,
  author={Yi, Ji Soo and Kang, Youn-Ah and Stasko, John and Jacko, J.A.},
  journal=TVCG, 
  title={Toward a Deeper Understanding of the Role of Interaction in Information Visualization}, 
  year={2007},
  volume={13},
  number={6},
  pages={1224--1231},
}

@article{yao2023tree,
  title={Tree of thoughts: Deliberate problem solving with large language models},
  author={Yao, Shunyu and Yu, Dian and Zhao, Jeffrey and Shafran, Izhak and Griffiths, Tom and Cao, Yuan and Narasimhan, Karthik},
  journal={NeurIPS},
  volume={36},
  pages={11809--11822},
  year={2023}
}

@article{pike2009science,
  title={The science of interaction},
  author={Pike, William A and Stasko, John and Chang, Remco and O'connell, Theresa A},
  journal={Information visualization},
  volume={8},
  number={4},
  pages={263--274},
  year={2009},
  publisher={SAGE Publications Sage UK: London, England}
}

@inproceedings{Zhu_multimodal_finance2025,
author = {Zhu, Fengbin and Li, Junfeng and Pan, Liangming and Wang, Wenjie and Feng, Fuli and Wang, Chao and Luan, Huanbo and Chua, Tat-Seng},
title = {Towards Temporal-Aware Multi-Modal Retrieval Augemented Generation in Finance},
year = {2025},
isbn = {9798400720352},
url = {https://doi.org/10.1145/3746027.3755723},
doi = {10.1145/3746027.3755723},
booktitle = ACMMM,
pages = {6289–6297},
numpages = {9},
location = {Dublin, Ireland},
series = {MM '25}
}

@inproceedings{Bin_GalleryGPT_2025,
author = {Bin, Yi and Shi, Wenhao and Ding, Yujuan and Hu, Zhiqiang and Wang, Zheng and Yang, Yang and Ng, See-Kiong and Shen, Heng Tao},
title = {GalleryGPT: Analyzing Paintings with Large Multimodal Models},
year = {2024},
isbn = {9798400706868},
url = {https://doi.org/10.1145/3664647.3681656},
doi = {10.1145/3664647.3681656},
booktitle = ACMMM,
pages = {7734–7743},
numpages = {10},
location = {Melbourne VIC, Australia},
series = {MM '24}
}

@inproceedings{Shuai_artRag_2025,
author = {Wang, Shuai and Najdenkoska, Ivona and Zhu, Hongyi and Rudinac, Stevan and Kackovic, Monika and Wijnberg, Nachoem and Worring, Marcel},
title = {ArtRAG: Retrieval-Augmented Generation with Structured Context for Visual Art Understanding},
year = {2025},
isbn = {9798400720352},
url = {https://doi.org/10.1145/3746027.3755673},
doi = {10.1145/3746027.3755673},
booktitle = ACMMM,
pages = {6700–6709},
numpages = {10},
location = {Dublin, Ireland},
series = {MM '25}
}

@inproceedings{Meng_GraphVideoAgent_2025,
author = {Chu, Meng and Li, Yicong and Chua, Tat-Seng},
title = {GraphVideoAgent: Enhancing Long-form Video Understanding with Entity Relation Graphs},
year = {2025},
isbn = {9798400720352},
url = {https://doi.org/10.1145/3746027.3755537},
doi = {10.1145/3746027.3755537},
booktitle = ACMMM,
pages = {4639–4648},
numpages = {10},
location = {Dublin, Ireland},
series = {MM '25}
}

@inproceedings{Rachabatuni_artChatBot_2024,
author = {Rachabatuni, Pavan Kartheek and Principi, Filippo and Mazzanti, Paolo and Bertini, Marco},
title = {Context-aware chatbot using MLLMs for Cultural Heritage},
year = {2024},
isbn = {9798400704123},
url = {https://doi.org/10.1145/3625468.3652193},
doi = {10.1145/3625468.3652193},
booktitle = {ACM Multimedia Systems Conference},
pages = {459–463},
numpages = {5},
location = {Bari, Italy},
series = {MMSys '24}
}

@inproceedings{Ma_reinforcement_retrieval_2024,
author = {Ma, Zhixin and Ngo, Chong Wah},
title = {Interactive Video Corpus Moment Retrieval using Reinforcement Learning},
year = {2022},
isbn = {9781450392037},
url = {https://doi.org/10.1145/3503161.3548277},
doi = {10.1145/3503161.3548277},
booktitle = ACMMM,
pages = {296–306},
numpages = {11},
location = {Lisboa, Portugal},
series = {MM '22}
}

@inproceedings{Khan_Castle_2025,
author = {Khan, Omar Shahbaz and Sharma, Ujjwal and Marcelino, Gon\c{c}alo and Duane, Aaron and Rudinac, Stevan and Worring, Marcel and J\'{o}nsson, Bj\"{o}rn \TH{}\'{o}r},
title = {Interactive Retrieval System for Multi-Stream Collections: multiXview at CASTLE 2025 Interactive Grand Challenge},
year = {2025},
isbn = {9798400720352},
url = {https://doi.org/10.1145/3746027.3760243},
doi = {10.1145/3746027.3760243},
booktitle = ACMMM,
pages = {14273–14279},
numpages = {7},
location = {Dublin, Ireland},
series = {MM '25}
}

@inproceedings{NII-UIT-VBS2026,
author = {Tran, Bao and Do, Tien and Ngo, Thanh Duc and Le, Duy-Dinh and Satoh, Shin’ichi},
title = {NII-UIT at VBS2026: Towards Effective Visual Question Answering for Interactive and Multimodal Video Retrieval},
year = {2026},
isbn = {978-981-95-6962-5},
url = {https://doi.org/10.1007/978-981-95-6963-2_26},
doi = {10.1007/978-981-95-6963-2_26},
booktitle = {MultiMedia Modeling: 32nd International Conference on Multimedia Modeling, MMM 2026, Prague, Czech Republic, January 29–31, 2026, Proceedings, Part IV},
pages = {238–244},
numpages = {7},
location = {Prague, Czech Republic}
}

@inproceedings{Prak_VBS2026,
author = {J\"{a}ckl, Bastian and Verner, Benjamin and Stroh, Michael and Kloda, Vojt\v{e}ch and Nagy, Ladislav and Deussen, Oliver and Keim, Daniel A. and Loko\v{c}, Jakub},
title = {PraK V4 at the Video Browser Showdown 2026},
year = {2026},
isbn = {978-981-95-6962-5},
url = {https://doi.org/10.1007/978-981-95-6963-2_25},
doi = {10.1007/978-981-95-6963-2_25},
booktitle = {MultiMedia Modeling: 32nd International Conference on Multimedia Modeling, MMM 2026, Prague, Czech Republic, January 29–31, 2026, Proceedings, Part IV},
pages = {230–237},
numpages = {8},
location = {Prague, Czech Republic}
}

@inproceedings{Exquisitor_VBS2026,
author = {Khan, Omar Shahbaz and Sharma, Ujjwal and Marcelino, Gon\c{c}alo and Rudinac, Stevan and J\'{o}nsson, Bj\"{o}rn \TH{}\'{o}r},
title = {Exquisitor at the Video Browser Showdown 2026: Temporal Queries Revisited},
year = {2026},
isbn = {978-981-95-6962-5},
url = {https://doi.org/10.1007/978-981-95-6963-2_27},
doi = {10.1007/978-981-95-6963-2_27},
booktitle = {MultiMedia Modeling: 32nd International Conference on Multimedia Modeling, MMM 2026, Prague, Czech Republic, January 29–31, 2026, Proceedings, Part IV},
pages = {245–251},
numpages = {7},
location = {Prague, Czech Republic}
}
